\documentclass[aps,pra,superscriptaddress,nofootinbib]{revtex4}
\pdfoutput=1
\usepackage[intlimits]{amsmath}
\usepackage{amsfonts}
\usepackage{psfrag}
\usepackage{array}
\usepackage{comment}
\usepackage{epsfig}
\usepackage{bm}
\usepackage{mathtools}
\DeclarePairedDelimiter{\ceil}{\lceil}{\rceil}
\usepackage{color}
\usepackage{times}
\usepackage{braket}
\usepackage{cancel}
\usepackage[colorlinks,bookmarks=false,citecolor=blue,linkcolor=red,urlcolor=blue]{hyperref}
\hypersetup{%
  colorlinks = true,
  linkcolor  = black
} 

\def\doi{http://dx.doi.org/}
\usepackage{tikz}
\usetikzlibrary{arrows}
\usetikzlibrary{decorations.pathreplacing}

\usepackage[colorlinks,bookmarks=false,citecolor=blue,linkcolor=red,urlcolor=blue]{hyperref}
\hypersetup{%
  colorlinks = true,
  linkcolor  = black
} 

\def\doi{http://dx.doi.org/}

\newcommand{\be}{\begin{equation}}
\newcommand{\ee}{\end{equation}}
\newcommand{\bea}{\begin{eqnarray}}
\newcommand{\eea}{\end{eqnarray}}
\newcommand{\nn}{\nonumber \\}

\advance\voffset by -1cm
\advance\hoffset by -0.5cm
\begin{document}

\title{Phase separation in the six-vertex model with a variety of boundary conditions}

\author{I. Lyberg}
\affiliation{Instituto Internacional de Fisica, UFRN,  Campos Universit\' ario, Lagoa Nova  59078-970 Natal, Brazil}
\author{V. Korepin}
\affiliation{C.N. Yang Institute for Theoretical Physics, Stony Brook University, Stony Brook, USA}
\author{G.A.P. Ribeiro}
\affiliation{Departamento de F\'{i}sica, Universidade Federal de S\~ao Carlos \\ S\~ao Carlos, SP 13565-905, Brazil}
\author{J. Viti} 
\affiliation{ECT \& Instituto Internacional de F\'isica, UFRN,  Campos Universit\' ario, Lagoa Nova  59078-970 Natal, Brazil}

\begin{abstract}
We present numerical results for the six-vertex model with a variety of boundary conditions. Adapting an algorithm proposed by Allison and
Reshetikhin~\cite{AR2005} for domain wall boundary conditions, we examine some modifications of these boundary conditions. To be precise, we discuss partial domain wall boundary conditions, reflecting ends and
half turn boundary conditions (domain wall boundary conditions with half turn symmetry).
\end{abstract}
\maketitle
\section{Introduction}
\label{intro}

The six-vertex model is a famous model in statistical mechanics~\cite{BAXTER}. In a nutshell, one can consider a square lattice and draw
arrows along its edges. The vertices must  however satisfy the so-called \textit{ice rule}:  each of them should always host  the same number of
incoming and outgoing arrows (for a square lattice such a number is obviously two). Summing over all the arrow configurations
that respect the ice rule with the proper Boltzmann weights turns out a complicated
combinatorial problem.
Nevertheless, the free energy in the thermodynamic limit with periodic boundary conditions  was exactly determined by Lieb in \cite{LIEB}; soon
afterwards  the same result was also recovered with free boundary conditions~\cite{WU}. In both cases the phase diagram is
controlled by a single parameter $\Delta$ that is a combination of the
Boltzmann weights $w$'s; see for instance Sec.~\ref{sec2} and in particular Eq.~\eqref{delta}.

Because of the constraint imposed by the ice rule, it was already noticed in~\cite{WU} that the physical
properties of the six-vertex model cannot be independent from the arrow configuration selected at the  boundary even
in the thermodynamic limit.
This observation motivated over the years the study of alternative but still
exactly solvable boundary conditions;
see for instance \cite{OWCZAREK,BATCHELOR} where the same free energy was found as earlier by Lieb.
A non-trivial demonstration of the decisive role played by the boundary in the six-vertex model  was possible only
after the introduction~\cite{K1982} of the so-called Domain Wall Boundary conditions (DWBC).
DWBC are special boundary conditions in which all the arrows in the upper and lower sides
of the square lattice are
incoming while those in the left-hand and right-hand sides are  outgoing. 
Due to the fact that the partition function of the model with DWBC can be expressed in terms of
a determinant \cite{IZERGIN,KOREPIN1992}, it was possible to analytically evaluate
the free-energy in the thermodynamic limit~\cite{KOREPIN2000,ZINNJUSTIN,BLEHER}. 
The six-vertex model has also been studied numerically \cite{BN1998,SZ2004,AR2005,LKV2017,KL2017}.
Numerical studies have showed that typical configurations exhibit phase separation~\cite{SZ2004, AR2005},
which was also observed in the study of domino tiling of the Aztec diamond \cite{PROPP1998}. 
For instance  when $|\Delta|\leq 1$  a central disordered region can coexist surrounded 
by four frozen corners.
The curves that in the thermodynamic limit separate the different phases are termed arctic curves.
When $\Delta = 0$ such a  curve was proven to be an ellipse~\cite{PROPP1998} that actually simplifies, for symmetric weights, into a circle dubbed the
\textit{arctic circle}.
Arctic curves were later determined with Bethe Ansatz methods for all  values of $\Delta$, where they are now known in parametric
form~\cite{PC2010, PKZJ, CS2016}.

Similarly, the six-vertex model with one reflecting end~\cite{T1998}, was also showed
to have a different thermodynamic free-energy compared
to the classical result by Lieb (at least in the disordered and ferromagnetic regimes~\cite{RK2015}).
The dependence of the physical properties of the six-vertex model 
on
the boundary conditions
was further elucidated in \cite{TAVARES}, underlining how these  can continuously vary
by
modifying the arrow orientation at the boundary. Arguably,  DWBC are then  only a particular case 
among those exactly solvable cases for which the six-vertex model can display phase separation
and configurations contain an arctic curve.
It is then a relevant question to understand how
phase coexistence is realized beyond such a paradigmatic example.  
For instance, analogous boundary conditions, especially related to alternating-sign matrix enumeration problems, have appeared in the
literature~\cite{KUPERBERG1996,KUPERBERG2002,FW2012} for which partition functions  could be expressed through suitable determinants or pfaffians.
Moreover the thermodynamic limit of the partition function has been calculated  for 
half-turn boundary conditions  when $w(a_1)=w(a_2)$ and $w(b_1)=w(b_2)$ (or DWBC with half turn symmetry, see~\cite{BL2017}) and partial domain wall boundary conditions in  ferromagnetic phase~\cite{BL2014}. In the former  case the extensive part of the free energy is the same as for DWBC.  

At present, the evidence of spatially separated phases, although expected,  and the precise form of the arctic curves there  remain  open problems. Some progress~\cite{DF2017} however has been made recently  employing the so-called tangent method introduced in~\cite{CS2016}.
In order to  comprehensively address these questions, we numerically analyze with a
Monte Carlo  algorithm proposed in~\cite{AR2005} by Allison and Reshetikhin,
the six-vertex model with a variety of  boundary conditions. Such a numerical method has been
showed in~\cite{LKV2017} to reproduce accurately  known analytic results for the six-vertex model with DWBC. Simulations in~\cite{LKV2017} were also compatible with those performed in~\cite{KL2017} that implemented a different algorithm and extensively checked it. 
It is then reasonable to expect that
a similar agreement should be found for closely analogous  boundary conditions. It must be stressed  that the Metropolis algorithm might be however impractical in the antiferromagnetic phase, where the system may be unable to thermalize. 

The paper is then organized as follows: In Sec.~\ref{sec2} we illustrate the numerical method  in the
context of the six-vertex model with DWBC. We present  numerical results for phase separation and density  profiles  with partial domain wall boundary conditions (in Sec.~\ref{sec_pdwbc}), one reflecting end (in Sec.~\ref{rdwbc}) and half turn boundary conditions (in Sec.~\ref{sec_ht}).
Finally, we summarize in Sec.~\ref{sec_con} our conclusions. Two appendices complete the paper. In Appendix~\ref{apa} we report results about the six-vertex model with DWBC at $\Delta=-2$ and $b=2a$ that are used to benchmark the reliability of the algorithm in the antiferromagnetic regime. In Appendix~\ref{apb} we show   convergence of the density profiles, obtained in Sec.~\ref{rdwbc} and Sec.~\ref{sec_ht}, to their exact conjectured value in the thermodynamic limit.

\section{Allison-Reshetikhin algorithm and density profiles}
\label{sec2}

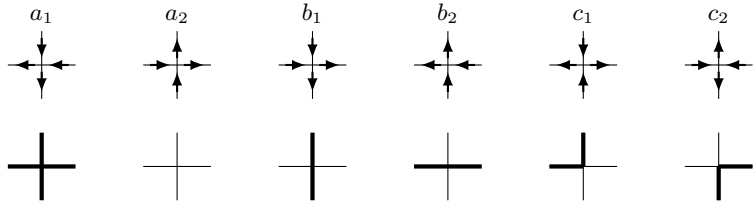
\begin{figure}[t]
\centering
\begin{tikzpicture}[scale=0.9]
\draw(-0.5,0)--(0.5,0);
\draw(0,-0.5)--(0,0.5);
\draw [thick, -latex] (-0.1,0.0) to (-0.4,0);
\draw [thick, -latex] (0.4,0.0) to (0.1,0);
\draw [thick, -latex] (0,0.4) to (0.0,0.1);
\draw [thick, -latex] (0.0,-0.1) to (0.0,-0.4);

\draw [ultra thick](-0.5,-1.5)--(0.5,-1.5);
\draw [ultra thick] (0,-2.0)--(0,-1.0);
\node[above] at (0,0.5) {$a_1$};

\begin{scope}[xshift=2.0cm]
\draw(-0.5,0)--(0.5,0);
\draw(0,-0.5)--(0,0.5);
\draw [thick, -latex] (-0.4,0.0) to (-0.1,0);
\draw [thick, -latex] (0.1,0.0) to (0.4,0);
\draw [thick, -latex] (0.0,0.1) to (0.0,0.4);
\draw [thick, -latex] (0,-0.4) to (0.0,-0.1);
\draw (-0.5,-1.5)--(0.5,-1.5);
\draw  (0,-2.0)--(0,-1.0);
\node[above] at (0,0.5) {$a_2$};
\end{scope}

\begin{scope}[xshift=4.0cm]
\draw(-0.5,0)--(0.5,0);
\draw(0,-0.5)--(0,0.5);
\draw [thick, -latex] (-0.4,0.0) to (-0.1,0);
\draw [thick, -latex] (0.1,0.0) to (0.4,0);
\draw [thick, -latex] (0.0,0.4) to (0.0,0.1);
\draw [thick, -latex] (0.0,-0.1) to (0.0,-0.4);
\draw (-0.5,-1.5)--(0.5,-1.5);
\draw [ultra thick] (0,-2.0)--(0,-1.0);
\node[above] at (0,0.5) {$b_1$};
\end{scope}

\begin{scope}[xshift=6.0cm]
\draw(-0.5,0)--(0.5,0);
\draw(0,-0.5)--(0,0.5);
\draw [thick, -latex] (-0.1,0.0) to (-0.4,0);
\draw [thick, -latex] (0.4,0.0) to (0.1,0);
\draw [thick, -latex] (0.0,0.1) to (0.0,0.4);
\draw [thick, -latex] (0.0,-0.4) to (0.0,-0.1);
\draw [ultra thick](-0.5,-1.5)--(0.5,-1.5);
\draw  (0,-2.0)--(0,-1.0);
\node[above] at (0,0.5) {$b_2$};
\end{scope}

\begin{scope}[xshift=8.0cm]
\draw(-0.5,0)--(0.5,0);
\draw(0,-0.5)--(0,0.5);
\draw [thick, -latex] (-0.1,0.0) to (-0.4,0);
\draw [thick, -latex] (0.1,0.0) to (0.4,0);
\draw [thick, -latex] (0.0,0.4) to (0.0,0.1);
\draw [thick, -latex] (0.0,-0.4) to (0.0,-0.1);
\draw (-0.5,-1.5)--(0.5,-1.5);
\draw  (0,-2.0)--(0,-1.0);
\draw [ultra thick](-0.5,-1.5)--(0.0,-1.5);
\draw [ultra thick] (-0.0,-1.0)--(0.0,-1.5);
\node[above] at (0,0.5) {$c_1$};
\end{scope}

\begin{scope}[xshift=10.0cm]
\draw(-0.5,0)--(0.5,0);
\draw(0,-0.5)--(0,0.5);
\draw [thick, -latex] (-0.4,0) to (-0.1,0);
\draw [thick, -latex] (0.4,0.0) to (0.1,0);
\draw [thick, -latex] (0.0,0.1) to (0.0,0.4);
\draw [thick, -latex] (0.0,-0.1) to (0.0,-0.4);
\draw (-0.5,-1.5)--(0.5,-1.5);
\draw  (0,-2.0)--(0,-1.0);
\draw [ultra thick](0.5,-1.5)--(0.0,-1.5);
\draw [ultra thick] (-0.0,-2.0)--(0.0,-1.5);
\node[above] at (0,0.5) {$c_2$};
\end{scope}

\end{tikzpicture}
\caption{The six vertices. From left to right: $a_1$, $a_2$, $b_1$, $b_2$, $c_1$, and $c_2$. The upper row shows the arrow representation, and the lower row shows the line representation.
An arrow pointing left or down is represented by a thick line.}
\label{fig:sixvertices}
\end{figure}


\textit{Metropolis algorithm.}~We start by reviewing the Allison-Reshetikhin (AR) algorithm~\cite{AR2005}.  The algorithm was originally formulated for the six-vertex model with DWBC  
but it is valid in principle   for any fixed boundary conditions. For simplicity in this section, we will exemplified  it in the case of DWBC.

\begin{figure}[t]
\centering
\begin{tikzpicture}[scale=1]
\draw[<->, >=latex](-0.5,0)--(3.5,0);
\draw[<->, >=latex](-0.5,1)--(3.5,1);
\draw[<->, >=latex](-0.5,2)--(3.5,2);
\draw[<->, >=latex](-0.5,3)--(3.5,3);
\draw[>-<, >=latex](0,-0.4)--(0,3.4);
\draw[>-<, >=latex](1,-0.4)--(1,3.4);
\draw[>-<, >=latex](2,-0.4)--(2,3.4);
\draw[>-<, >=latex](3,-0.4)--(3,3.4);
\draw(0,-0.5)--(0,3.4);
\draw(1,-0.5)--(1,3.4);
\draw(2,-0.5)--(2,3.4);
\draw(3,-0.5)--(3,3.4);
\draw[red, dashed,->,-latex](-0.5,-0.5)--(-0.5,3.8);
\draw[red, dashed](3.5,-0.5)--(3.5,3.5);
\draw[red, dashed](-0.5,3.5)--(3.5,3.5);
\draw[red, dashed,->,-latex](-0.5,-0.5)--(3.8,-0.5);
\node[below] at (3.8,-0.5) {$x$};
\node[left] at (-0.5,3.8) {$y$};
\node[above] at (1.75,3.5) {$N=4$};
\draw[very thick](0,3.5)--(0,3);
\draw[very thick](-0.5,3.0)--(0.0,3.0);
\draw[very thick](-0.5,2.0)--(1.0,2.0);
\draw[very thick](1.0,2.0)--(1.0,3.5);
\draw[very thick](-0.5,1.0)--(1.0,1.0);
\draw[very thick](1.0,1.0)--(1.0,2.0);
\draw[very thick](1.0,2.0)--(2.0,2.0);
\draw[very thick](2.0,2.0)--(2.0,3.5);
\draw[very thick](-0.5,0.0)--(2.0,0.0);
\draw[very thick](2.0,0.0)--(2.0,1.0);
\draw[very thick](2.0,1.0)--(3.0,1.0);
\draw[very thick](3.0,1.0)--(3.0,3.5);
\node[red] at (1,2) {$\bullet$};
\node[below] at (1.5,2) {$(x,y)$};

\node[above] () at (0,-1.0) {$0$};
\node[above] () at (1,-1.0) {$1$};
\node[above] () at (2,-1.0) {$2$};
\node[above] () at (3,-1.0) {$3$};
\node[right] () at (-1,0.0) {$0$};
\node[right] () at (-1,1.0) {$1$};
\node[right] () at (-1,2.0) {$2$};
\node[right] () at (-1,3.0) {$3$};

\node at (0,-1.2){(a)};

\begin{scope}[xshift=7cm]
\draw[<->, >=latex](-0.5,0)--(5.5,0);
\draw[<->, >=latex](-0.5,1)--(5.5,1);
\draw[<->, >=latex](-0.5,2)--(5.5,2);
\draw[<->, >=latex](-0.5,3)--(5.5,3);
\draw[>-, >=latex](0,-0.4)--(0,3.5);
\draw[>-, >=latex](1,-0.4)--(1,3.5);
\draw[>-, >=latex](2,-0.4)--(2,3.5);
\draw[>-, >=latex](3,-0.4)--(3,3.5);
\draw[>-, >=latex](4,-0.4)--(4,3.5);
\draw[>-, >=latex](5,-0.4)--(5,3.5);

\draw(0,-0.5)--(0,3.5);
\draw(1,-0.5)--(1,3.5);
\draw(2,-0.5)--(2,3.5);
\draw(3,-0.5)--(3,3.5);
\draw(4,-0.5)--(4,3.5);
\draw(5,-0.5)--(5,3.5);
\draw[very thick](-0.5,3.0)--(1.0,3);
\draw[very thick](1.0,3.0)--(1.0,3.5);
\draw[very thick](-0.5,2.0)--(1.0,2.0);
\draw[very thick](1.0,2.0)--(1.0,3.0);
\draw[very thick](1.0,3.0)--(2.0,3.0);
\draw[very thick](2.0,3.0)--(2.0,3.5);
\draw[very thick](-0.5,1.0)--(2.0,1.0);
\draw[very thick](2.0,1.0)--(2.0,2.0);
\draw[very thick](2.0,2.0)--(4.0,2.0);
\draw[very thick](4.0,2.0)--(4.0,3.5);
\draw[very thick](-0.5,0.0)--(4.0,0.0);
\draw[very thick](4.0,0.0)--(4.0,2.0);
\draw[very thick](4.0,2.0)--(5.0,2.0);
\draw[very thick](5.0,2.0)--(5.0,3.5);
\node[red] at (1,3) {$\bullet$};
\node at (0,-1.2){(b)};
\end{scope}
\end{tikzpicture}
\caption{(a)~The six-vertex model with DWBC with $N=4$. The picture also shows an allowed state. In such a  state, there are two vertices flippable up
at $(x,y)=(2,0)$ and at $(3,1)$, one vertex flippable down at $(2,1)$,
and one vertex which is flippable both up and down; the red vertex at $(1,2)$. (b)~Partial domain wall boundary conditions. Here $M=6$ and $N=4$. The figure also shows 
one state of the six-vertex model with pDWBC. In this state, there are two
vertices flippable up, one vertex flippable down,
and one vertex flippable both left and down, the red vertex.
In addition, there is one vertex flippable right (only).}
\label{dwbcfig}
\end{figure}
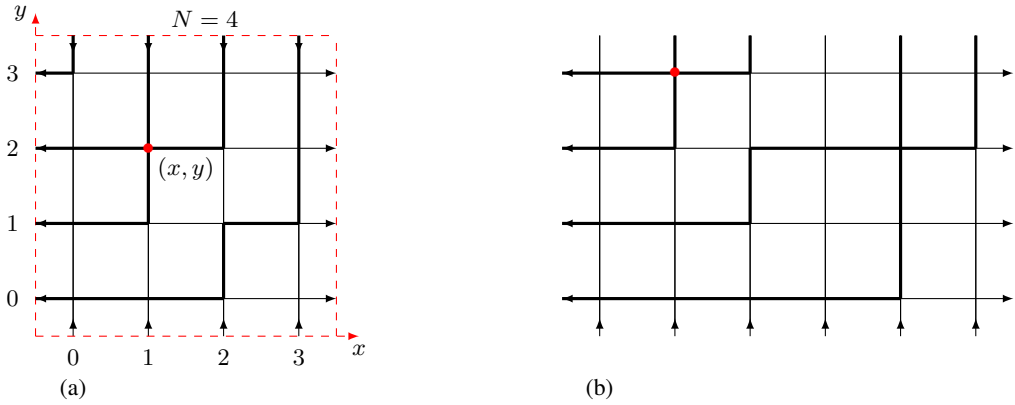

First consider the six-vertex model on a square lattice of dimensions $N\times N$. The lattice vertices $v=(x,y)$ have integer coordinates with 
$0\leq x\leq (N-1)$ and $0\leq y \leq (N-1)$ as in Fig.~\ref{dwbcfig}a and we  denote the six possible vertex states  $s_v$ by $a_1$, $a_2$, $b_1$, $b_2$, $c_1$, and $c_2$;
see Fig. \ref{fig:sixvertices}.
Each state comes with a Boltzmann
weight $w(s_v)$. Four vertices  lying at the corners of  a square of side-length equal to the
lattice spacing define a plaquette. The state of the plaquette $S$ is obviously defined by the states of the
vertices that form it and the weight of the plaquette that contains $v$ we will denote by $W_{v}(S)$. Such a weight is the product of all
the vertex weights in the plaquette, $W_v(S)=\prod_{u\in\text{plaquette}}w(s_u)$. This is a slight abuse of notation: There are four plaquettes that contain $v$, but it should be clear from the context to which we will be referring to.
Furthermore, in vertex $a_1$,
we interpret the figure in the lower row as two curves (thick lines) that do not cross.
The first comes in from above and goes out on the left, and the second comes in from the right and goes out below. All the other five vertices can be interpreted analogously.

DWBC for the six-vertex model are then defined as follows: all the arrows on the upper and lower vertical edges of the square lattice point inward  whereas those on the leftmost and rightmost horizontal edges point outwards. It is easy to realize that  the curves can enter the lattice only from  above and leave the lattice only from the left side. Moreover, each allowed state with DWBC can be represented as a set of $N$ curves; an example
is showed in Fig.~\ref{dwbcfig}a.
Two different curves may meet at a vertex, but they can not cross each other or share an edge. It is not difficult to show that~\cite{AR2005},  due to conservation laws, it is possible to choose  $w(a_1)=w(a_2)=a$, $w(b_1)=w(b_2)=b$ and $w(c_1)=w(c_2)=c$. Furthermore the partition function depends, apart a constant factor, only on $a/c$  and $b /c$, leaving only two independent parameters.   It is customary  to characterize the phase diagram with DWBC borrowing the same terminology that is used for periodic boundary conditions~\cite{BAXTER}; we then introduce the parameter
\begin{equation}
\label{delta}
\Delta=\frac{a^2+b^2-c^2}{2ab},
\end{equation}
and refer to $|\Delta|\leq 1$ as the disordered regime, while $\Delta<-1$ and the $\Delta>1$ are the antiferromagnetic and ferromagnetic regimes respectively. 

The AR algorithm has two Monte Carlo (MC) moves;
called up flip and down flip; see Fig. \ref{pdwbcfig}a for an example of a down flip.
An up flip at the vertex $(x,y)$  takes the plaquette 
$\{(x,y),(x,y+1),(x-1,y+1),(x-1,y)\}$ from a state $S$ to a new state $S'$.
It is easy to see that 
there are sixteen distinct plaquette states $S$ under the six-vertex rule where such a move is possible. These sixteen states together have 
up to nine distinct
weights. A given vertex may be flippable up only, down only, up and down, or not flippable. As an example, see Fig. \ref{dwbcfig}a.
The computer code  has three lists, each one containing the vertices flippable in one of three different ways. The lists will be called
$L_u$, $L_d$ and $L_{ud}$. We can formulate the algorithm as follows.
First, randomly search among all vertices in the lattice, until a vertex contained in one of the three lists is found.
A flip will then be attempted, and if the attempt is successful, then all three lists will be updated near the
vertex where the flip occurred. The probabilities of attempted flips are as follows. We first consider a vertex
$v$ flippable up only. Such a vertex is flipped with the probability 
\be P_u = \frac{W_v(S')}{R},
\label{eq1}
\ee
where $S'$ is the state of the plaquette after the flip, and $R$ is a positive constant,
large enough that no probability is greater than 1. The constant
$R$ is arbitrary and it only affects the thermalization time, as discussed in~\cite{AR2005}.
A vertex flippable down only is treated similarly, and we call $P_d$ the probability of a down flip.
When a vertex $v$ is flippable both up and down, then
a flip will occur with the sum of the two probabilities
\be P_{ud} = \frac{W_v(S'_1)+W_v(S'_2)}{R},
\label{eq2}
\ee
where $S_1'$ and $S_2'$ are the two possible states after a flip (pertaining to two different plaquettes); we say that  $S'_{1}$
will be the new state of the plaquette after a flip up and $S'_{2}$ the state (of a different plaquette) after a flip down.
If it is decided that there should be a flip, then it has to be decided which way it should be. With
probability
\be P_{u|ud} = \frac{W_v(S'_1)}{W_v(S'_1)+W_v(S'_2)}
\label{eq3}
\ee
the flip will be up to the state $S'_1$ (analogously down to $S'_2$). The above conditions satisfy detailed balance and ensure ergodicity.

 We define a MC sweep as $N^2$ executed vertex flips. In units of MC sweeps, the integrated autocorrelation time for the density
of type-c vertices also grows sub-linearly with the system size. To ensure the system has thermalized we proceed as in~\cite{LKV2017} and  compare the arctic curve obtained from the numerics for sufficiently large $N$ ($N=250$) with its exact thermodynamic limit~\cite{PC2010, PKZJ}. For the cases where the arctic curve is not known we monitor whether the average external boundary
of typical configurations does not change. We always check that for the same values of parameters used in the simulation, the exact arctic curve with DWBC is recovered. In particular, we notice that for  large and negative values of $\Delta$, typical configurations contain a central antiferromagnetic  phase, that can slower significantly thermalization and especially when $a\not= b$. For this reason we do not consider cases where $\Delta<-2$ with our algorithm, see Appendix~\ref{apa} about this point.  After thermalization is reached, the MC code averages any observable $O(x,y)$ 
over a large number of typical states obtaining its equilibrium value that is denoted by $\langle O(x,y)\rangle$.

\textit{Density profiles.~}The observables we use to study the behaviour of the model are the density profiles; see \cite{KL2017} and \cite{LKV2017} for a detailed study 
of various density profiles in the case of DWBC.
One density is the density of the curves which separate the level sets; we call this density $\rho$ \cite{LKV2017}. This density was used in \cite{LKV2017}, and it is defined as
\begin{equation}
\rho(x,y) = \left\{
\begin{array}{rl}
1 & \text{if the vertex at ~} (x,y) \text{~is~} a_1 ,\\
0 & \text{if the vertex at ~} (x,y) \text{~is~} a_2,\\
1/2 & \text{if the vertex at ~} (x,y) \text{~is neither~} a_1 \text{~nor~} a_2 .
\end{array} \right.
\end{equation} 
Another density, which will be used in this paper, is the difference of the density vertices of type $c_1$ and vertices of type $c_2$. This density~\cite{KL2017} is called $\delta\rho_c$
\begin{equation}
\delta\rho_c(x,y) = \left\{
\begin{array}{rl}
1 & \text{if the vertex at ~} (x,y) \text{~is~} c_1 ,\\
-1 & \text{if the vertex at ~} (x,y) \text{~is~} c_2,\\
0 & \text{if the vertex at ~} (x,y) \text{~is neither~} c_1 \text{~nor~} c_2 .
\end{array} \right.
\end{equation} 
In addition to the above densities on vertices, we have defined the density on edges.
Edges are labelled by their midpoint coordinates $(x,y)$: for vertical edges  $y$  is half-integer and $x$ is integer; for horizontal edges $x$ is half-integer and $y$ is integer.  
The densities on vertical and horizontal edges 
are called $\rho_v$ and $\rho_h$. The former is defined as
\begin{equation}
\rho_v(x,y) = \left\{
\begin{array}{rl}
1 & \text{if the arrow on edge ~} (x,y) \text{~points down} ,\\
0 & \text{if the arrow on edge ~} (x,y) \text{~points up}.
\end{array} \right.
\end{equation} 
The density $\rho_h$ can be defined analogously, but we will not discuss it in this paper.
\section{Six-vertex model with Partial Domain Wall Boundary Conditions}
\label{sec_pdwbc}

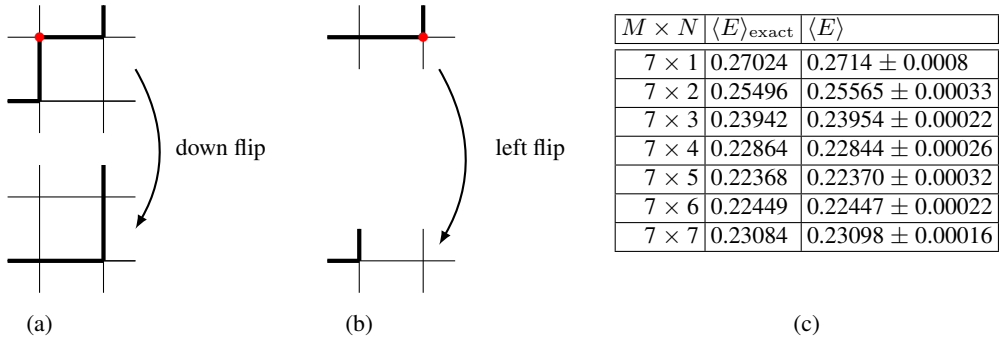
\begin{figure}[t]
\centering
\begin{tikzpicture}[scale=0.85]
\draw (-0.5,0)--(1.5,0);
\draw (-0.5,1)--(1.5,1);
\draw (-0.5,0)--(1.5,0);
\draw (0,-0.5)--(0,1.5);
\draw (1,-0.5)--(1,1.5);
\draw[ultra thick] (-0.5,0)--(1,0);
\draw[ultra thick] (1,0)--(1,1.5);
\draw (-0.5,0+2.5)--(1.5,0+2.5);
\draw (-0.5,1+2.5)--(1.5,1+2.5);
\draw (-0.5,0+2.5)--(1.5,0+2.5);
\draw (0,-0.5+2.5)--(0,1.5+2.5);
\draw (1,-0.5+2.5)--(1,1.5+2.5);
\draw[ultra thick] (-0.5,2.5)--(0,2.5);
\draw[ultra thick] (0,2.5)--(0,3.5);
\draw[ultra thick] (0,3.5)--(1,3.5);
\draw[ultra thick] (1,3.5)--(1,4);
\draw[thick, -latex](-0.5+2,0.5+2.5) to [bend left](-0.5+2,0.5);
\node[red] at (0,3.5){$\bullet$};
\node[right] at (2,1.75) {\rm{down flip}};
\node at (0,-1){(a)};
\begin{scope}[xshift=5cm]
\node at (0,-1){(b)};
 \draw(-0.5,0.0+3.5)--(1.5,0+3.5);
\draw(0,-0.5+3.5)--(0,0.5+3.5);
\draw(1,-0.5+3.5)--(1,0.5+3.5);
\draw[ultra thick](-0.5,0.0+3.5)--(1.0,0+3.5);
\draw[ultra thick](1.0,0+3.5)--(1,0.5+3.5);
\draw(-0.5,0.0)--(1.5,0);
\draw(0,-0.5)--(0,0.5);
\draw(1,-0.5)--(1,0.5);
\draw[ultra thick](-0.5,0.0)--(0,0);
\draw[ultra thick](0,0)--(0,0.5);
\draw[thick, -latex](-0.5+1.75,0.5+2.5) to [bend left](-0.5+1.75,0.25);
\node[right] at (2,1.75) {\rm{left flip}};
\node[red] at (1,3.5){$\bullet$};
\end{scope}
\node at (12,-1){(c)};
\node at (12,2){\begin{tabular}{|r|l|l|}
\hline
$M \times N$ &  $\langle E\rangle_{\rm{exact}}$ & $\langle E\rangle$  \\
\hline \hline 
7 $\times$ 1  & 0.27024 & 0.2714 $\pm$ 0.0008 \\ 
\hline
7 $\times$ 2  & 0.25496 & 0.25565 $\pm$ 0.00033 \\ 
\hline
7 $\times$ 3  & 0.23942 & 0.23954 $\pm$ 0.00022 \\ 
\hline
7 $\times$ 4  & 0.22864 & 0.22844 $\pm$ 0.00026 \\ 
\hline
7 $\times$ 5  & 0.22368 & 0.22370 $\pm$ 0.00032 \\ 
\hline
7 $\times$ 6  & 0.22449 & 0.22447 $\pm$ 0.00022 \\ 
\hline
7 $\times$ 7  & 0.23084 & 0.23098 $\pm$ 0.00016 \\ 
\hline
\end{tabular}};
\end{tikzpicture}

\caption{(a)~The down flip. In the upper plaquette, we say that the upper left corner is flippable down. The lower plaquette shows the state after the flip, which is called $S'$.
The states of the four vertices on the left are, beginning from the upper left and going clockwise, $a_2$, $b_1$, $c_1$, and $b_2$.
Therefore $W_v(S')=w(a_2)w(b_1)w(b_2)w(c_1)$. In the case of DWBC, $w(a_1)=w(a_2)=a$, $w(b_1)=w(b_2)=b$ and $w(c_1)=w(c_2)=c$, therefore $W_v(S')=ab^2c$ in this case. 
(b)~The left flip. This move is possible with pDWBC when $M>N$. It can only happen on the upper boundary.
In the pair of vertices on top, we say that the right vertex is flippable left. On the bottom is seen the state on the upper boundary after the left flip, which is called $T'$. 
The states of the two vertices after the flip are, left to right, $c_1$ and $a_2$. Taking as for pDWBC,
$w(c_1)=w(c_2)=c$, we get $Z_v(T')=w(a_2)c$. (c)~
Mean vertex energy computed from exact enumeration of the six-vertex configurations with pDWBC and from the MC algorithm.
The weights are $w(a_1)=w(a_2)=3/5$, $w(b_1)=w(b_2)=4/5$, $c=1$. The agreement is excellent.}
\label{pdwbcfig}
\end{figure}

\textit{Model and algorithm.~}Partial Domain Wall Boundary conditions (pDWBC) are defined on a rectangular lattice of dimensions $M\times N$, where we assume $M\geq N$.
One of the long sides has free boundary condition,
and the other three sides have the same boundary conditions as the lattice with DWBC; see Fig. \ref{dwbcfig}b, where also an allowed state is showed. 
With pDWBC the six-vertex model has three independent parameters~\cite{BL2014}, contrary to DWBC. The weights $w(c_1)$ and $w(c_2)$ can be still taken equal to $c$ and without loss of generality $c=1$. However it turns out that $w(a_1)=ae^{-\eta}$, $w(a_2)=ae^{\eta}$, $w(b_1)=be^{-\eta}$, $w(b_2)=be^{\eta}$, thus introducing the new parameter $\eta$. 

With pDWBC, there are two more MC  moves on the upper boundary, which we shall call left flip
and right flip. Now seven lists have to be kept. Apart from the three lists mentioned above, these lists
are of vertices flippable left only, right only, left and right, and left and down. 
The four new lists will be called $L_l$, $L_r$, $L_{lr}$ and $L_{ld}$. Examples of vertices flippable
right only and of vertices flippable both left and down are seen in Fig. \ref{dwbcfig}b; notice that there are no vertices flippable right and down.

A left flip (see  Fig.~\ref{pdwbcfig}b) is an operation
on two neighbouring vertices on the upper boundary. There are four different vertex pairs  where a left flip is possible,
and these have among them up to four distinct weights. 
We will describe the case where a  left flip or a  down flip are both possible at the same vertex.
We assume that the vertex $v$ is such a vertex. At $v$, we can flip down to a new plaquette state $S'$ or left
to a new pair of vertices on the upper boundary $T'$. A flip (left or down) will occur with probability
\be P_{ld} = \frac{W_v(S')+Z_v(T')}{R},
\label{eq4}
\ee
where $Z_v(T')$ is the weight of the state $T'$ after a left flip. Just as in Sec. \ref{sec2}, $R$ is a sufficiently large positive constant.
 If it is decided that there will be a flip, then the flip will be left
  with probability $P_{l|ld}$ and down with probability $P_{d|ld}$ where
\be P_{l|ld} = \frac{Z_v(T')}{W_v(S')+Z_v(T')},\quad P_{d|ld} = \frac{W_v(S')}{W_v(S')+Z_v(T')}.
\label{eq5}
\ee
Note how different Eqs. (\ref{eq4}), (\ref{eq5}) are from Eqs. (\ref{eq2})
and (\ref{eq3}). In  Eq.~(\ref{eq2}), the weights $W_v(S'_1)$ and $W_v(S'_2)$ are both quartic monomials
in the variables $w(a_1)$, $w(a_2)$, $w(b_1)$, $w(b_2)$, and $c$; whereas in  Eq.~(\ref{eq4}), $W_v(S')$ is a quartic monomial 
and $Z_v(T')$ is a quadratic monomial. It can be showed 
in complete analogy with the case of DWBC that such an algorithm satisfies
detailed balance and it is ergodic. To confirm  its validity we performed
a comparison for small rectangles with a brute force sum over all the
vertex configurations satisfying the boundary conditions and the ice-rule.
In particular in the table on  Fig.~\ref{pdwbcfig}c we compared
the  mean vertex energy obtained by direct enumeration $\langle E\rangle_{\rm{exact}}$ with that obtained from the MC, $\langle E\rangle$. The energy  
$E(s_v)$ of a vertex $v$ in a state $s_v$ is defined  obviously as
$w(s_v)\equiv e^{-E(s_v)}$, being $w(s_v)$ the Boltzmann weight of $s_v$. As it can be seen, the accuracy of the MC is excellent and on the same footing as for DWBC; see the last row in the table in Fig~\ref{pdwbcfig}c. 
\newline
\newline

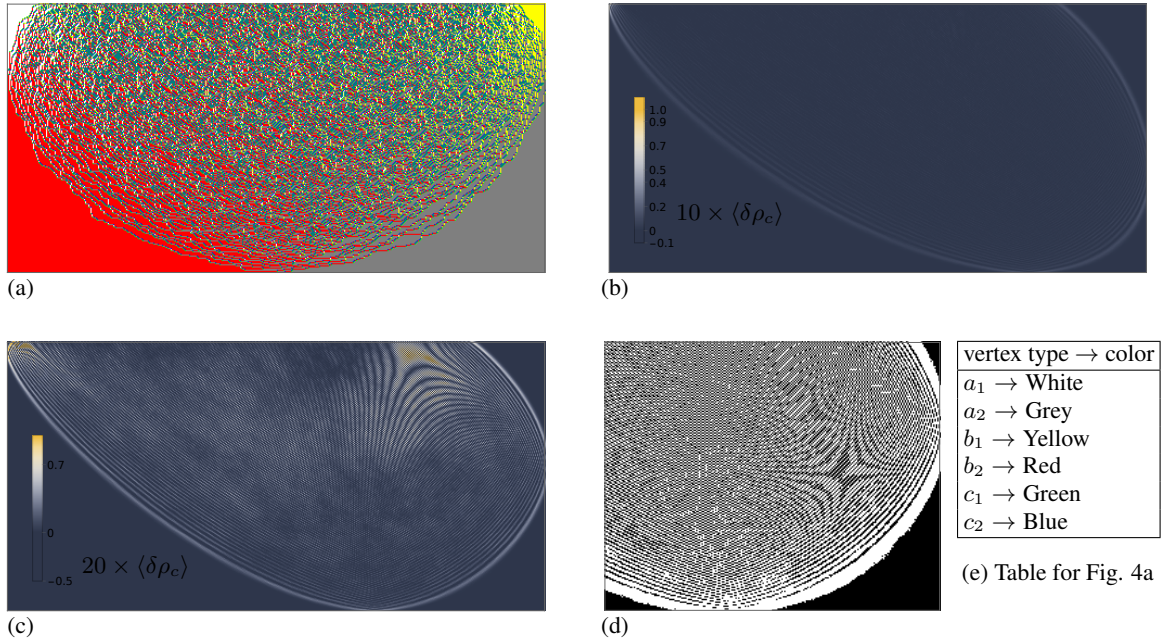
\begin{figure}
\centering
\begin{tikzpicture}
\node at (0,0){\includegraphics[width=0.45\textwidth]{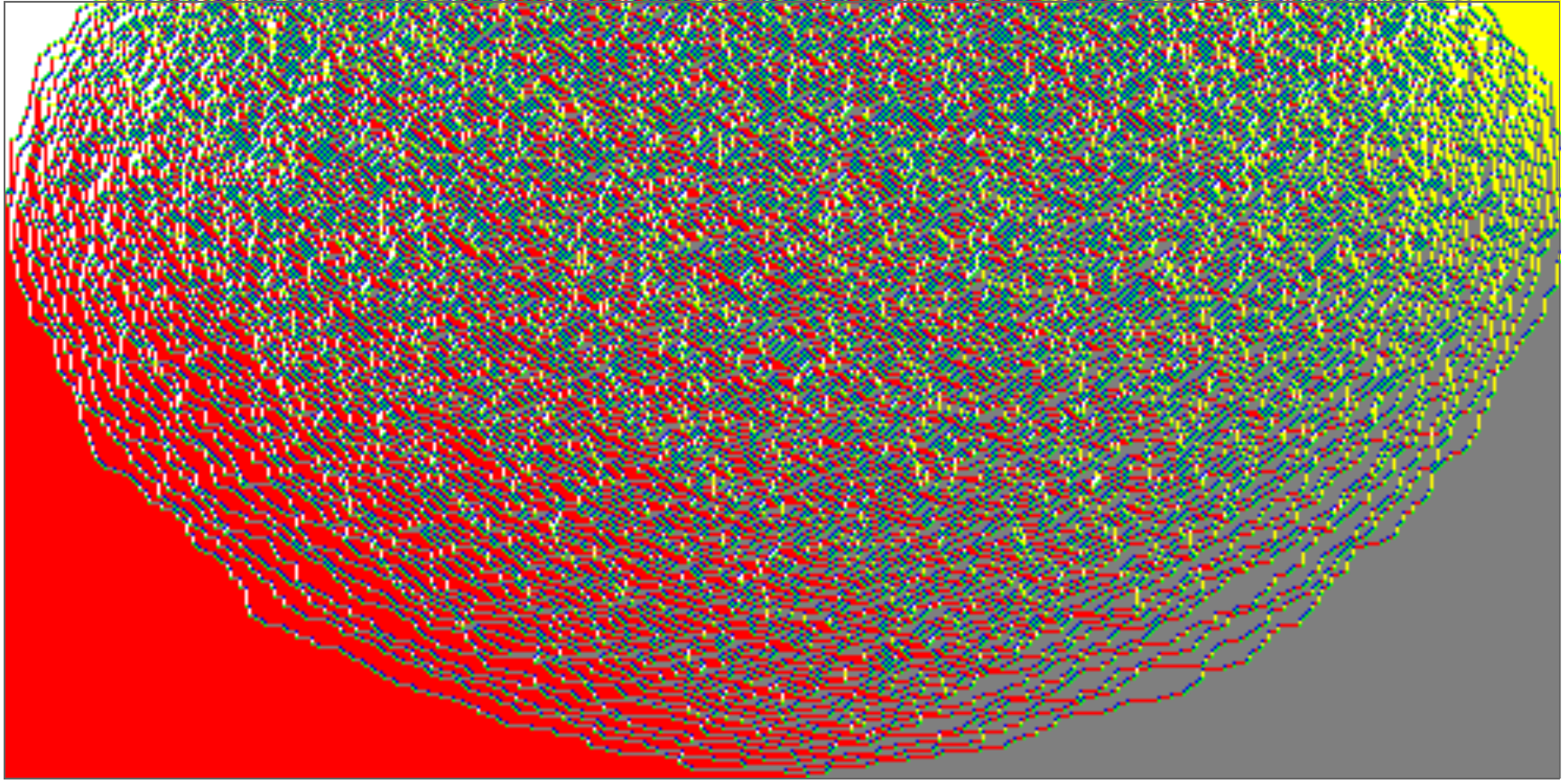}};
\node at (-3.4,-2){(a)};
\node at (4.5,-2){(b)};
\node at (-3.4,-6.5){(c)};
\node at (4.5,-6.5){(d)};
\node[right] at (9,-5.8){(e) Table for Fig. 4a};
\node at (8,0){\includegraphics[width=0.45\textwidth]{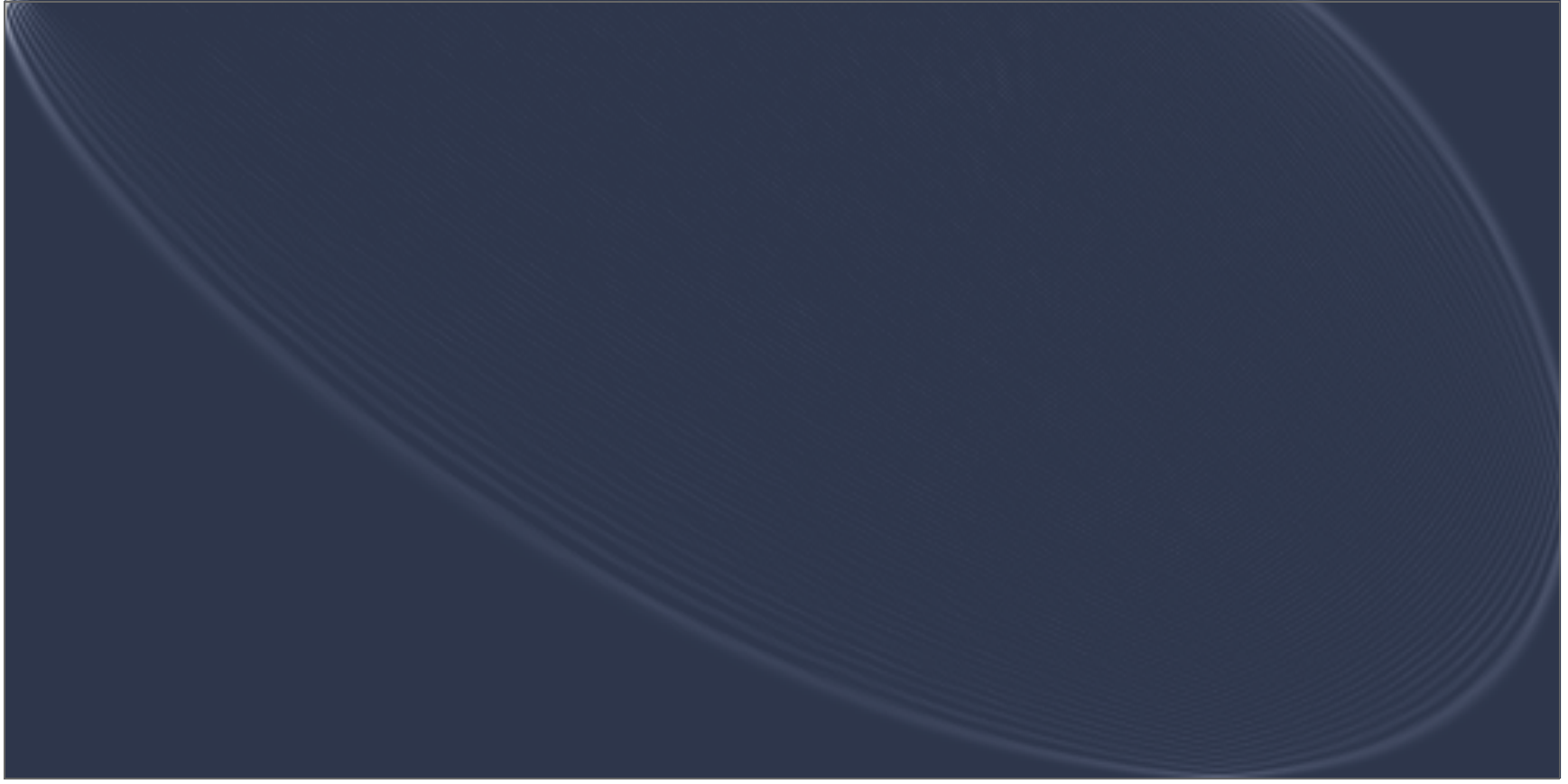}};
\node[right] at (5.2,-1){$10\times\langle\delta\rho_c\rangle$};
\node at (5,-0.5){\includegraphics[height=2.2cm]{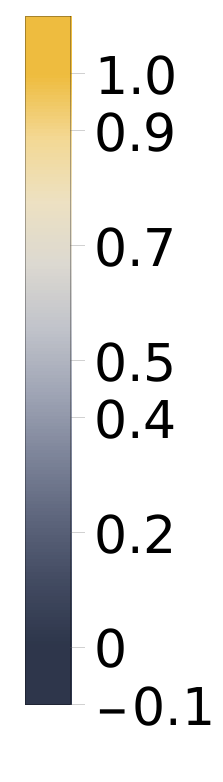}};
\node at (0,-4.5){\includegraphics[width=0.45\textwidth]{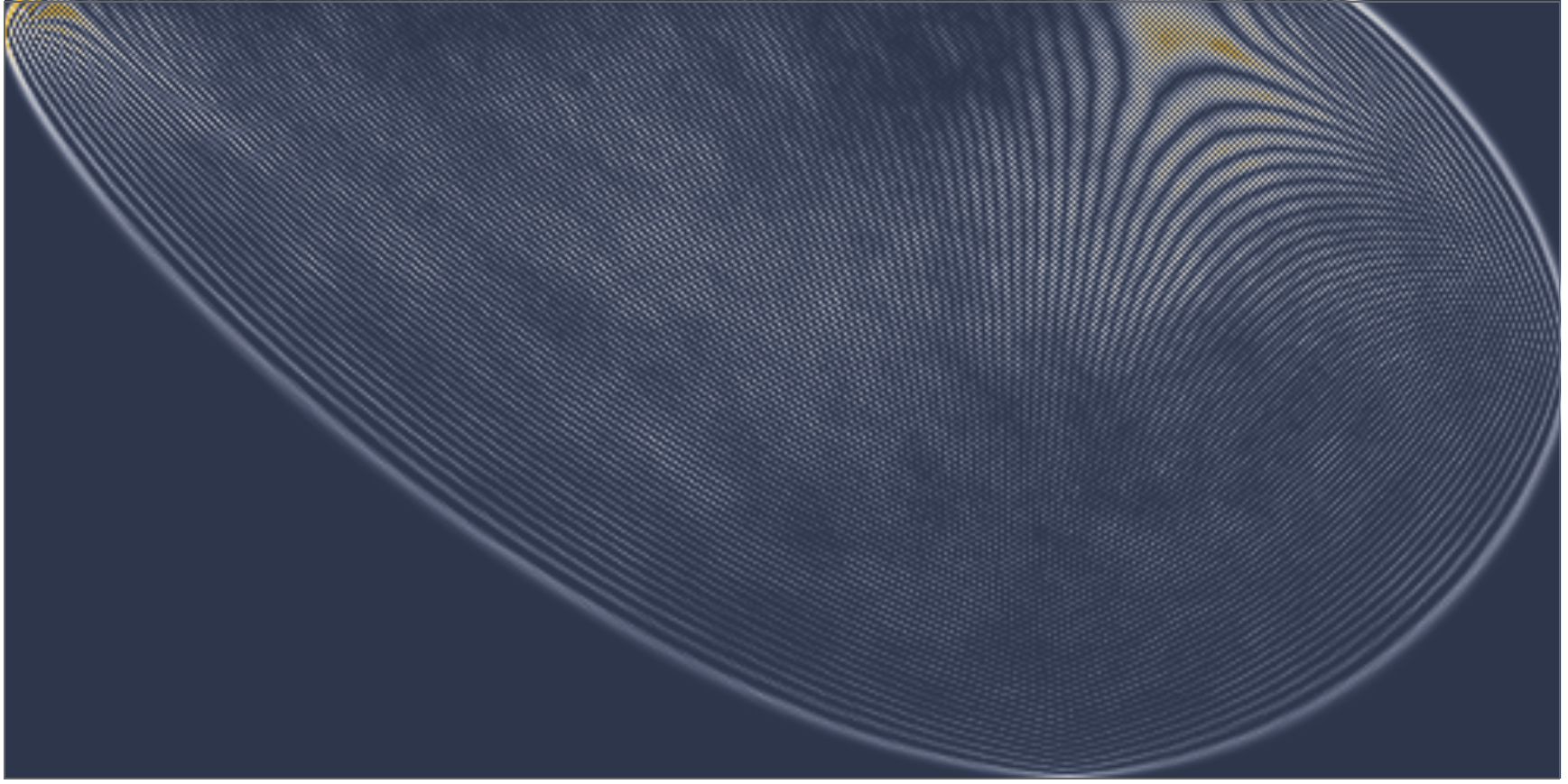}};
\node at (-3,-5){\includegraphics[height=2.2cm]{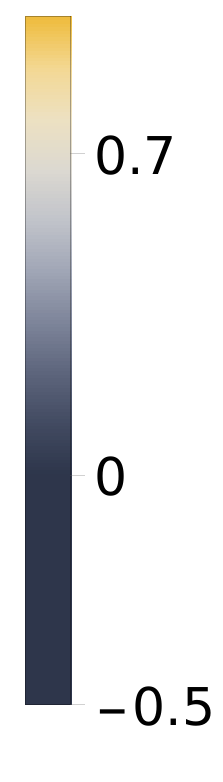}};
\node[right] at (-2.7,-5.7){$20\times\langle\delta\rho_c\rangle$};
\node at (6.6, -4.5){\includegraphics[height=3.6cm]{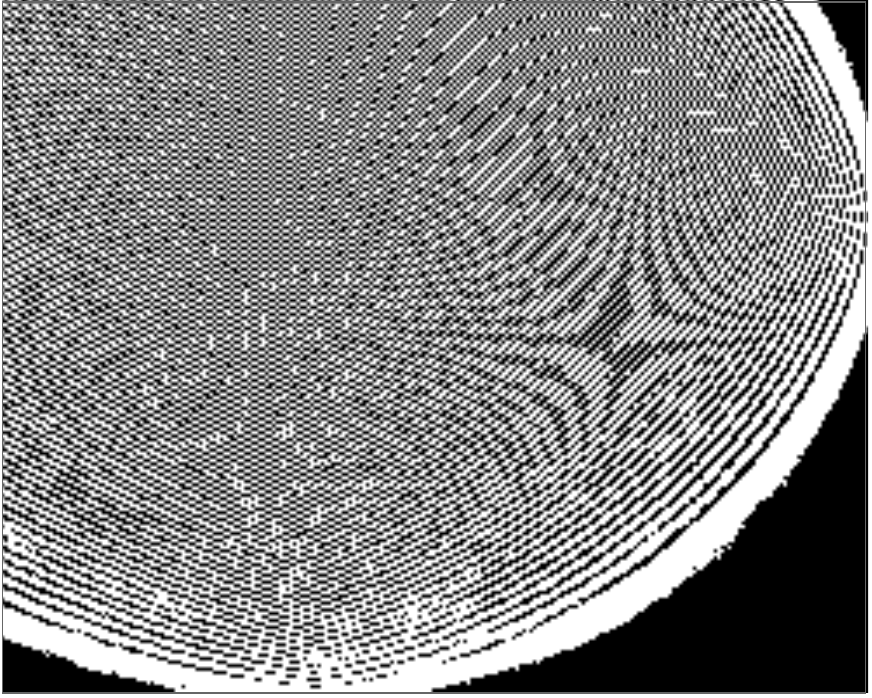}};
\node at (10.4,-4){\begin{tabular}{|l|}
\hline
\text{vertex type }$\rightarrow$ \text{color}\\
\hline
$a_1\rightarrow\text{White}$  \\
$a_2\rightarrow\text{Grey}$  \\
$b_1\rightarrow\text{Yellow}$ \\
$b_2\rightarrow\text{Red}$ \\
$c_1\rightarrow\text{Green}$ \\
$c_2\rightarrow\text{Blue}$\\
\hline
\end{tabular}};
\end{tikzpicture}
\caption{Partial domain wall
boundary conditions. The simulations
are performed on a rectangle with $500\times 250$ vertices. (a) shows a typical configuration for $\Delta=0$, $a=b$ and $c=1$; (b) and (c) are density plot of $\langle\delta\rho_c\rangle$ 
for $\Delta=0$ and $b=2a$, $\Delta=-2$, $b=2a$. Finally (d) is a magnification of the right half of (c) that shows better the emergence of a saddle-point-like feature in the density 
$\langle \delta\rho_c\rangle$; 
vertices are coloured black if  $\langle \delta \rho_c\rangle\leq 0$ and white otherwise.}
\label{fig_PDWBC1}
\end{figure}
\textit{Numerical Results.~}
We restrict ourselves to the case $w(a_1)=w(a_2)=a$, $w(b_1)=w(b_2)=b$, $w(c_1)=w(c_2)=c=1$, i.e. $\eta=0$.
These restrictions are however not essential for the algorithm to work correctly. The average $\langle \cdot \rangle$ is taken
over thermalized MC configurations obtained through the algorithm described
earlier.   In Fig.~\ref{fig_PDWBC1}a we show a typical configuration in a $2N\times N$ rectangle with $N=250$ at $\Delta=0$, $a=b$. Vertices of different types are coloured differently, 
according to the in table in Fig.~\ref{fig_PDWBC1}e. We observe four frozen corners. The upper left and right corners have states $a_1$ and $b_1$, respectively, and 
the lower left and right corners have states $b_2$ and $a_1$, respectively.
The frozen corners are then the same as with DWBC. Furthermore it clearly emerges a disordered central region. In Fig.~\ref{fig_PDWBC1}b 
and Fig.~\ref{fig_PDWBC1}c
we show the average value  of $\delta\rho_c$,  defined in Sec.~\ref{sec2}. The
simulation parameters are: $b=2a$, $\Delta=0$; $b=2a$, $\Delta=-2$ and the lattice is a $2N\times N$ rectangle with $N=250$. In Fig.~\ref{fig_PDWBC1}b and Fig.~\ref{fig_PDWBC1}c every vertex $v=(x,y)$ 
is represented by one pixel; the corresponding value of $\langle\delta\rho_c\rangle$  can be read off from the bar legend. We will call this representation of numerical data, a  density plot.

Although only rigorously defined for $N\rightarrow\infty$, we can  identify  the arctic curve from the density plots.
The average density $\langle\delta\rho_c\rangle$ is indeed  zero in the frozen regions but can fluctuate inside the disordered phase. 
Similar evidence of phase separation can be found for any $\Delta\leq 1$, also when the aspect ratio of
the rectangle is different from here. Unfortunately, we are
not aware of any exact result concerning phase separation in the six-vertex with
pDWBC but certainly our numerical results directly confirm the existence of a piecewise differentiable arctic curve for any $\Delta\leq 1$.

Another interesting observation concerns the antiferromagnetic regime
where $\Delta<-1$: Fig.~\ref{fig_PDWBC1}c is an
example for $b=2a$ and $\Delta=-2$ on the same rectangular lattice. For such values of the parameters, the presence of vertices of type $c_1$ or $c_2$ is energetically favoured.
We can observe indeed the formation of two 
antiferromagnetic phases inside the disordered inhomogeneous domain
bounded by the arctic curve. 
Such  antiferromagnetic
 phases contain a chequerboard pattern of alternating $c_1$ and $c_2$ vertices, together with possible defects. One can spot a larger antiferromagnetic domain on the right and a smaller one near the top left corner. For $\Delta\rightarrow-\infty$ the two  phases should coincide with the antiferromagnetic ground states where no defect is present.  
Finally, in Fig.~\ref{fig_PDWBC1}d we show a magnification of bottom right corner of Fig.~\ref{fig_PDWBC1}c, where the vertices $(x,y)$ are now coloured in black if $\langle\delta\rho_c\rangle\leq0$ and white otherwise. As  noticed in~\cite{KL2017} for the case of DWBC, we observe also here the presence of saddle-point-like features emerging in the average density. 
\section{Six-vertex model with one Reflecting End}
\label{rdwbc}

\begin{figure}[t]
\centering
\vspace*{0.5cm}
\begin{tikzpicture}[scale=1]

\draw[<->, >=latex](-0.5,0)--(5.5,0);
\draw[<->, >=latex](-0.5,1)--(5.5,1);
\draw[<->, >=latex](-0.5,2)--(5.5,2);
\draw[>-, >=latex](0,-0.4)--(0,2.5);
\draw[>-, >=latex](1,-0.4)--(1,2.5);
\draw[>-, >=latex](2,-0.4)--(2,2.5);
\draw[>-, >=latex](3,-0.4)--(3,2.5);
\draw[>-, >=latex](4,-0.4)--(4,2.5);
\draw[>-, >=latex](5,-0.4)--(5,2.5);

\draw(0,-0.5)--(0,2.5);
\draw(1,-0.5)--(1,2.5);
\draw(2,-0.5)--(2,2.5);
\draw(3,-0.5)--(3,2.5);
\draw(4,-0.5)--(4,2.5);
\draw(5,-0.5)--(5,2.5);

\node[left] at (-0.5,0) {$\mu$};
\node[left] at (-0.5,1) {$\mu$};
\node[left] at (-0.5,2) {$\mu$};

\draw[very thick](-0.5,2.0)--(1.0,2.0);
\draw[very thick](1.0,2.0)--(1.0,2.5);
\draw[very thick](-0.5,1.0)--(1.0,1.0);
\draw[very thick](1.0,1.0)--(1.0,2.0);
\draw[very thick](1.0,2.0)--(2.0,2.0);
\draw[very thick](2.0,2.0)--(2.0,2.5);
\draw[very thick](-0.5,0.0)--(4.0,0.0);
\draw[very thick](4.0,0.0)--(4.0,2.0);
\draw[very thick](4.0,2.0)--(5.0,2.0);
\draw[very thick](5.0,2.0)--(5.0,2.5);

\node[blue] at (0.5,3) {$\bullet$};
\node[blue, above] at (0.5,3) {$k_-(\lambda,\zeta)$};
\node[below] at (0,-0.5) {$\lambda$};
\node[below] at (1,-0.5) {$-\lambda$};
\node[blue] at (2.5,3) {$\bullet$};
\node[blue, above] at (2.5,3) {$k_+(\lambda,\zeta)$};
\node[below] at (2,-0.5) {$\lambda$};
\node[below] at (3,-0.5) {$-\lambda$};
\node[blue] at (4.5,3) {$\bullet$};
\node[blue, above] at (4.5,3) {$k_-(\lambda,\zeta)$};
\node[below] at (4,-0.5) {$\lambda$};
\node[below] at (5,-0.5) {$-\lambda$};
\draw  plot [smooth, tension=2] coordinates { (0.0,2.5) (0.5,3.0) (1.0,2.5)};
\draw  plot [smooth, tension=2] coordinates { (2.0,2.5) (2.5,3.0) (3.0,2.5)};
\draw  plot [smooth, tension=2] coordinates { (4.0,2.5) (4.5,3.0) (5.0,2.5)};
\node at (0,-1.3) {(a)};

\begin{scope}[xshift=8cm, yshift=-0.6cm, scale=0.8]
\node at (0,-1) {(b)};
\draw[latex'-latex',thick] (-1.0,2.5)--(-1.0,4.5);
\node[left] (s) at (-1.0,3.5) {$s$};
\draw[->,-latex, thick](-0.5,0)--(5.5,0);
\draw(-0.5,1)--(5.5,1);
\draw(-0.5,2)--(5.5,2);
\draw(-0.5,3)--(5.5,3);
\draw(-0.5,4)--(5.5,4);
\draw(-0.5,5)--(5.5,5);
\node at (0,0){$\bullet$};
\node[above] at (0, 5.5) {$y$};
\node[right] at (5.5, 0) {$x$};
\draw[dashed, blue](-0.5,-0.5)--(-0.5,5.5);
\draw[dashed, blue](-0.5+6,-0.5)--(-0.5+6,5.5);
\draw[dashed, blue](-0.5,-0.5)--(-0.5+6,-0.5);
\draw[dashed, blue](-0.5,-0.5+6)--(-0.5+6,-0.5+6);
\draw[red, thick](-0.5,3)--(3,-0.5);
\node[above, red] at (0.5, 2) {$r_s$};
\draw[red, thick, dashed](-0.5,3+2)--(3+2,-0.5+2-2);
\node[above] at (2.5,5.5) {$2N=6$};
\draw[dashed](0,4.5)--(-1,4.5);
\draw[thick](0,-0.5)--(0,2.5);
\draw[->,-latex, thick](0,2.5)--(0,5.5);
\draw(1,-0.5)--(1,1.5);
\draw(1,1.5)--(1,5.5);
\draw(2,-0.5)--(2,0.5);
\draw(2,0.7)--(2,5.5);
\draw(3,-0.5)--(3,5.5);
\draw(4,-0.5)--(4,5.5);
\draw(5,-0.5)--(5,5.5);
\draw[dashed](-1.0,1.5)--(1.0,1.5);
\draw[dashed](-1.0,2.5)--(0.0,2.5);
\draw[latex'-latex',thick] (-1.0,1.5)--(-1.0,2.5);
\node[left] (t) at (-1,2.0) {$x$};
\node[right] at (1,1.5) {$e\equiv(x,y)$};
\node[blue] at (1,1.5) {$\bullet$};
\node[blue] at (0,2.5) {$\bullet$};
\node[blue] at (2,0.5) {$\bullet$};
\node[blue] at (3,-0.5) {$\bullet$};
\end{scope}
\end{tikzpicture}
\caption{(a)~Domain wall boundary conditions with one reflective end. Here $N=3$;
it is also showed an allowed state, depicted by thick curves. (b)~The path
along which the density on vertical edges was measured in reference \cite{LKV2017} is the thick red line $r_s$.
The distance between the line and the edge closest to the main diagonal (red dashed line) is called $s$.  
Clearly $0\leq x \leq 2N-2-s$. In this figure, $2N=6$, $s=2$, and $x=1$.}
\label{rdwbcfig}
\end{figure}
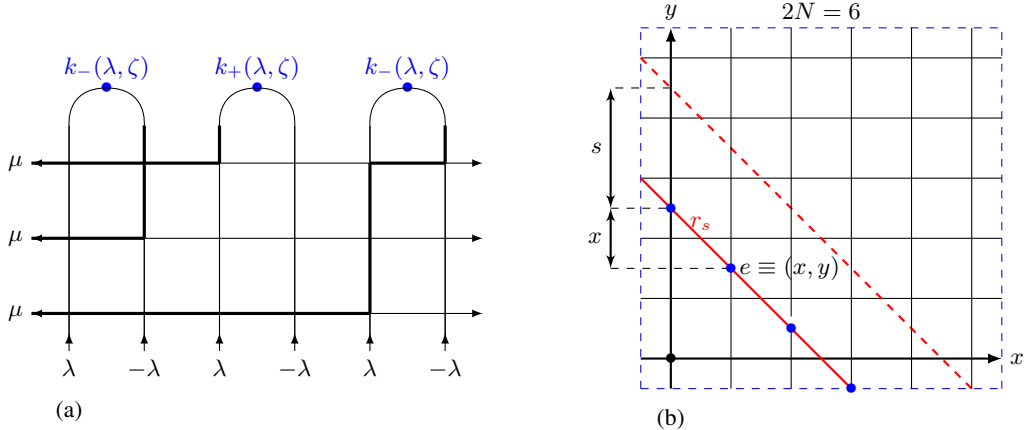

\textit{Model and algorithm.~}Domain Wall Boundary Conditions with one Reflective End (RE) are defined on a rectangular lattice of size $2N\times N$. The boundary conditions are similar to pDWBC, 
but with two important differences as we will discuss. 
Firstly, the vertices on the upper boundary are divided into $N$ pairs, where each pair consists of two nearest neighbours. In each pair, the two upper arrows have opposite directions; 
an example of an allowed configuration is shown
in Fig. \ref{rdwbcfig}a. In this case then, the upper boundary vertices
that are flippable right  can only have even $x$ coordinate and the ones flippable left can only have odd $x$ coordinate.

Moreover, in the model that is Bethe Ansatz solvable one introduces two spectral parameters $\lambda$ and $\mu$, pictorially running
on each vertical and horizontal line of the lattice as in  Fig.~\ref{rdwbcfig}a. The model can be solved for arbitrary complex values of $\lambda$ and $\mu$. We will restrict  to the case $\mu=0$ for simplicity, although a model with $\mu\not=0$ could be simulated as well. The six-vertex model with one reflecting end has then three vertex weights that,  for instance~\cite{RK2015}, in the disordered regime can be chosen ($\gamma<\lambda<\pi-\gamma$)
\bea a(\lambda) & = & \sin{(\gamma-\lambda)} \nn
b(\lambda) & = & \sin{(\gamma+\lambda)} \nn
c(\lambda) & = & \sin{2\gamma},
\label{abc}
\eea
where $0<\gamma <\pi/2$ and clearly, see Eq.~\eqref{delta}, $\Delta = -\cos{2\gamma}$. On even columns (see Figure \ref{rdwbcfig}a) $w(a_1)=w(a_2)=a(\lambda)$, $w(b_1)=w(b_2)=b(\lambda)$ and
$w(c_1)=w(c_2)=c$. However, because of the reflection  at the upper boundary~\cite{Skly}, on odd columns, the sign associated with the parameter $\lambda$ is reversed. Thus there
$w(a_1)=w(a_2)=b(\lambda)$, $w(b_1)=w(b_2)=a(\lambda)$. 
As for the case of DWBC, when $\mu = 0$, there are actually only two parameters in the phase diagram: $a/c$ and $b/c$. We can indeed divide all the entries in the $R$-matrix of the six-vertex model~\cite{RES_lect} by $c$ without spoiling the Yang-Baxter equations. Thus without loss of generality we assume $c=1$ in our simulations.
Finally, the model contains
two additional weights associated to the $U$-turn vertex, see Fig.~\ref{rdwbcfig}a: they are the diagonal entries of the reflection
matrix~\cite{Skly}. The weights are called $k_{\pm}(\lambda,\zeta)$ where $\zeta\in\mathbb C$ is the boundary parameter~\cite{RK2015} and they explicitly read  in the disordered regime
\begin{equation}
k_+(\lambda,\zeta)=\frac{\sin(\zeta+\lambda+\gamma)}{\sin(\zeta)},\quad k_{-}(\lambda,\zeta)=\frac{\sin(\zeta-\lambda-\gamma)}{\sin(\zeta)}.
\end{equation}
For $\zeta=\pi/2$, $k_{+}=k_{-}$ for any $\lambda$ and we are free to set the weights
 to 1, since the reflection equation is homogeneous~\cite{Skly}. Different $U$-turn weights can be easily implemented but we will not consider them here. The model is defined analogously~\cite{T1998} in the antiferromagnetic regime $\Delta<-1$ and the choice $k_+=k_-$ corresponds to $\zeta=i\pi/2$ there. To our best knowledge~\cite{RK2015} the thermodynamic limit of the partition functions it is not known in the antiferromagnetic regime. 
 
 To check again the validity of the algorithm we compare against the mean
vertex energy obtained from exact enumeration. For
instance for a $8\times 4$ rectangle with $a=3/5$, $b=4/5$ and $c=1$  we found 
$\langle E\rangle_{\rm{exact}}=0.224847\dots$
and $\langle E\rangle=0.224261\pm 0.00025$ from MC; the results
are compatible within the error bar.
 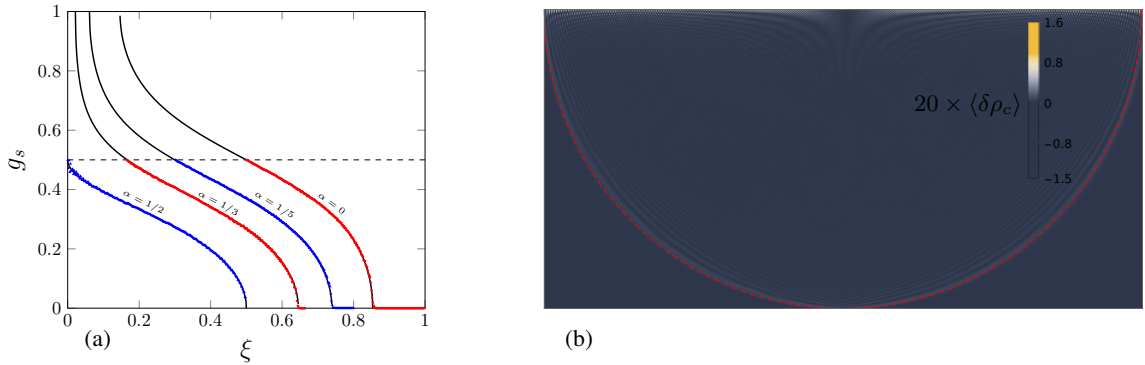
\begin{figure}[t]
\begin{center}
\begin{tikzpicture}[scale=0.8]\node at (-10,0.1) {\includegraphics[width=.35\textwidth]{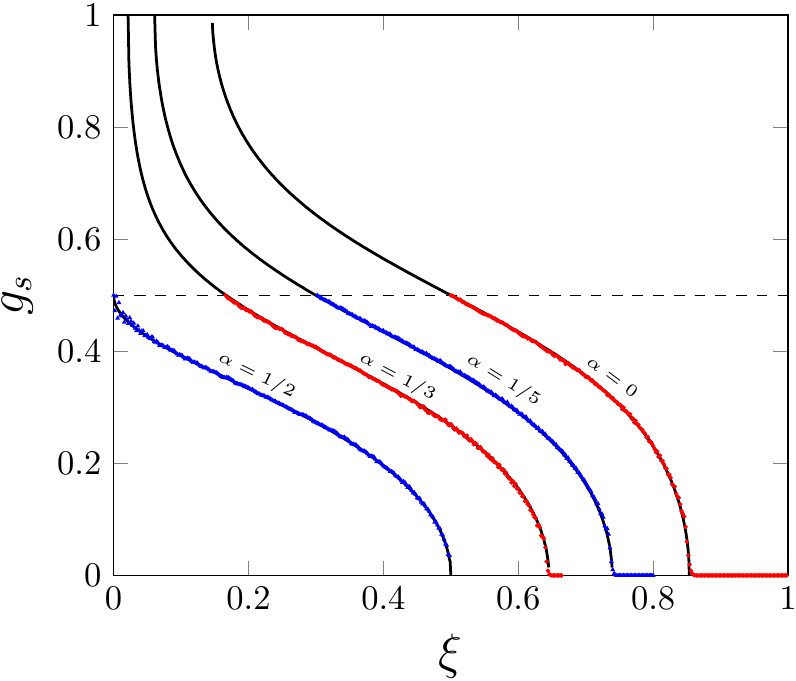}};
\node at (-12,-2.5){(a)};
\node at (-11.5+7.5,-2.5){(b)};
\node at (0.4,0.5){\includegraphics[width=0.5\textwidth]{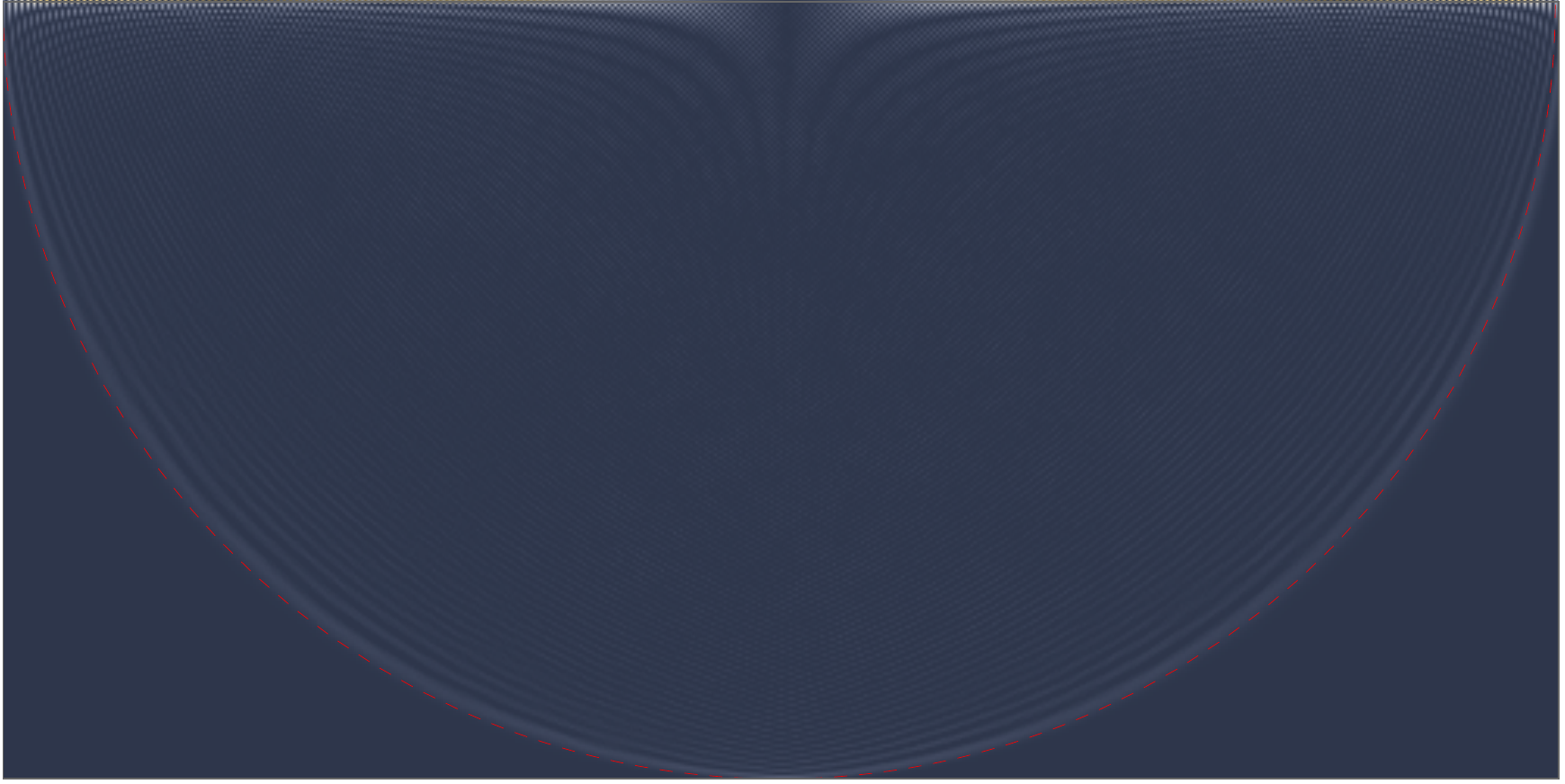}};
\node at (3.8,1.4){\includegraphics[height=2.4cm]{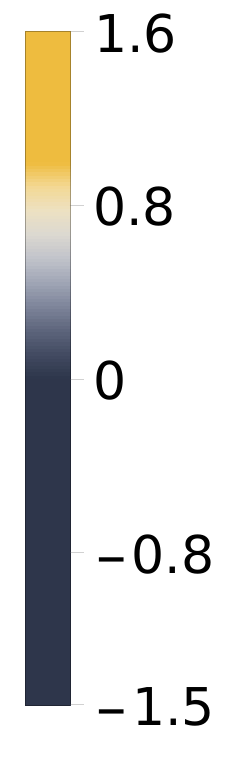}};
\node[left] at (3.5,1.4){$20\times\langle\delta\rho_c\rangle$};
\end{tikzpicture}
\caption{(a)~The  plot  shows a comparison between the numerical values of $g_s(x)$ obtained from the MC and the conjectured exact expression \eqref{conj} valid for $N\rightarrow\infty$. For a definition of  $\xi$ and $\alpha$ see the discussion in the text. Here $N=250$, $a=b$ and $\Delta=0$. (b)~Density plot of $\langle\delta\rho_c\rangle$ with $N=250$, $a=b$ and $\Delta=0$; a  red dashed semicircle, inscribed into a $2N\times N$ rectangle, has been superimposed.}
\label{vertical}
\end{center}
\end{figure}
\newline
\newline
\textit{  The arctic curve and the density profiles for $a=b$~} We use the MC algorithm to simulate the six-vertex model with with a RE for $\mu=0$, and equal $U$-turn weights (i.e. when the reflection matrix~\cite{Skly} is proportional to the identity matrix). We consider $N=250$ and we will show results for the representative cases $a=b$ and $\Delta=0$ and $b=2a$, $\Delta=-2$.  In particular,
Fig.~\ref{vertical}b shows the density plot of the average value $\langle\delta\rho_c\rangle$
estimated from the MC algorithm when $a=b$, $\Delta=0$ and $N=250$.  
The frozen lower left and right corners are characterized by $\langle\delta\rho_c\rangle=0$ as discussed in Sec.~\ref{sec_pdwbc} and a disordered region appears in the center bounded by an arctic curve. 
Such a curve is numerically found to be 
a semicircle: see again Fig.~\ref{vertical}b where also a red dashed semicircle inscribed into a $2N\times N$ rectangle is superimposed to the density plot. Keeping the constraint $a=b$, in all the cases we examined, we 
notice (see Fig.~\ref{fig_dwbcm2}b in Appendix~\ref{apa} for another example with $\Delta=-2$) that the limiting shape for the six-vertex model with a RE converges to the lower half of the exact curve~\cite{PC2010, PKZJ} calculated with DWBC. Moreover when $\Delta=0$, we also found that the average vertical edge density with RE converges  for sufficiently large
 $N$ to the same quantity computed on the lower half of a $2N\times 2N$ lattice with DWBC. More precisely, consider a  $2N\times 2N$ square whose lower half is a $2N\times N$ rectangle; on such a square draw the thick red  line $r_s$ showed
 on  Fig. \ref{rdwbcfig}b. According to the definitions settled in Sec. \ref{sec2}, we denote the midpoint edge coordinates by $e\equiv(x,y)$ where
$x$ is an integer and $y$ is half integer. The equation for  $r_s$  is then $\ceil{y}=-x+(2N-s-1)$ for $1\leq\ceil{y}\leq 2N-1-s$ and we consider $s$ for simplicity  in the interval $0\leq s\leq 2N-2$, i.e. $r_s$ is never above the dashed red line
in Fig.~\ref{rdwbcfig}b. 
Let $g_s(x)$ the average vertical edge density measured along $r_s$ on the lower rectangular lattice $2N\times N$ with a RE, i.e.
\be
g_s(x)=\langle \rho_v(x,y) \rangle|_{(x,y)\in r_s}^{\rm{RE}}.
\label{eq7}
\ee
Of course, the same quantity  is naturally defined also on the whole $2N\times 2N$ lattice with DWBC. We call therefore $f_s(x)$ the function $f_s(x)=\langle \rho_v(x,y) \rangle|_{(x,y)\in r_s}^{\rm{DW}}$. It can be proved, see for instance~\cite{ADSV2016}, that for any value of $a$ and $b$, the  limit $N\rightarrow\infty$ of
$f_{2N\alpha}(2N\xi)$ ($s\equiv 2N\alpha+O(1)$ for large $N$) exists and it is given by the elementary function
\begin{equation}
\label{asympt}
 \lim_{N\rightarrow\infty}f_{2N\alpha}(2N\xi)=\frac{1}{\pi}\arccos\left(\frac{\frac{b\alpha}{a}+\frac{a(2\xi+\alpha-1)}{b}}{\sqrt{1-(2\xi-1)^2}}\right)
\end{equation}
with $\xi\in[0,1]$ and $\alpha\in[0,1]$. 
When $a=b$ and $\Delta=0$, numerical data for RE, see Appendix B also, are compatible with~\eqref{asympt};  we conjecture then
 \begin{equation}
 \label{conj}
  \lim_{N\rightarrow\infty}g_{2N\alpha}(2N\xi)=\frac{1}{\pi}\arccos\left(\frac{2(\alpha+\xi)-1}{\sqrt{1-(2\xi-1)^2}}\right),
 \end{equation}
for $\xi\in[\max\{\frac{1}{2}-\alpha,0\},1-\alpha]$ and $\alpha\in[0,1]$. The plot in Fig.~\ref{vertical}a is a numerical verification  of the conjecture \eqref{conj}. The data obtained from the MC for
 the vertical edge density $g_{2N\alpha}(2N\xi)$ with $N=250$ are plotted for different values of $\alpha$ (alias $s$) within the allowed domain for $\xi$ (alias $x$). The black thick curves are instead obtained from \eqref{conj} and are indistinguishable from the numerical points already when $N=250$. The error bar is invisible on such a scale (see also Appendix~\ref{apb}).

\begin{figure}[t]
\centering
\hspace*{0.6cm}\begin{tikzpicture}
\draw[<-, >=latex](-0.5,0)--(3.5,0);
\draw[<-, >=latex](-0.5,1)--(3.5,1);
\draw[<-, >=latex](-0.5,2)--(3.5,2);
\draw[<-, >=latex](-0.5,3)--(3.5,3);
\draw[-, >=latex](0,-0.4)--(0,3.5);
\draw[-, >=latex](1,-0.4)--(1,3.5);
\draw[-, >=latex](2,-0.4)--(2,3.5);
\draw[-, >=latex](3,-0.4)--(3,3.5);
\draw[very thick](1,-0.5)--(1,3.5);
\draw[very thick](3,-0.5)--(3,3.5);
\draw[very thick](-0.5,0)--(3.5,0);
\draw[very thick](-0.5,1)--(3.5,1);
\draw[very thick](-0.5,2)--(3.5,2);
\draw[very thick](-0.5,3)--(3.5,3);
\draw  plot [smooth, tension=2] coordinates { (0.0,2.5+1) (0.5,3.0+1) (1.0,2.5+1)};
\draw  plot [smooth, tension=2] coordinates { (2.0,2.5+1) (2.5,3.0+1) (3.0,2.5+1)};
\node at (9.8,1.5){\includegraphics[width=0.5\textwidth]{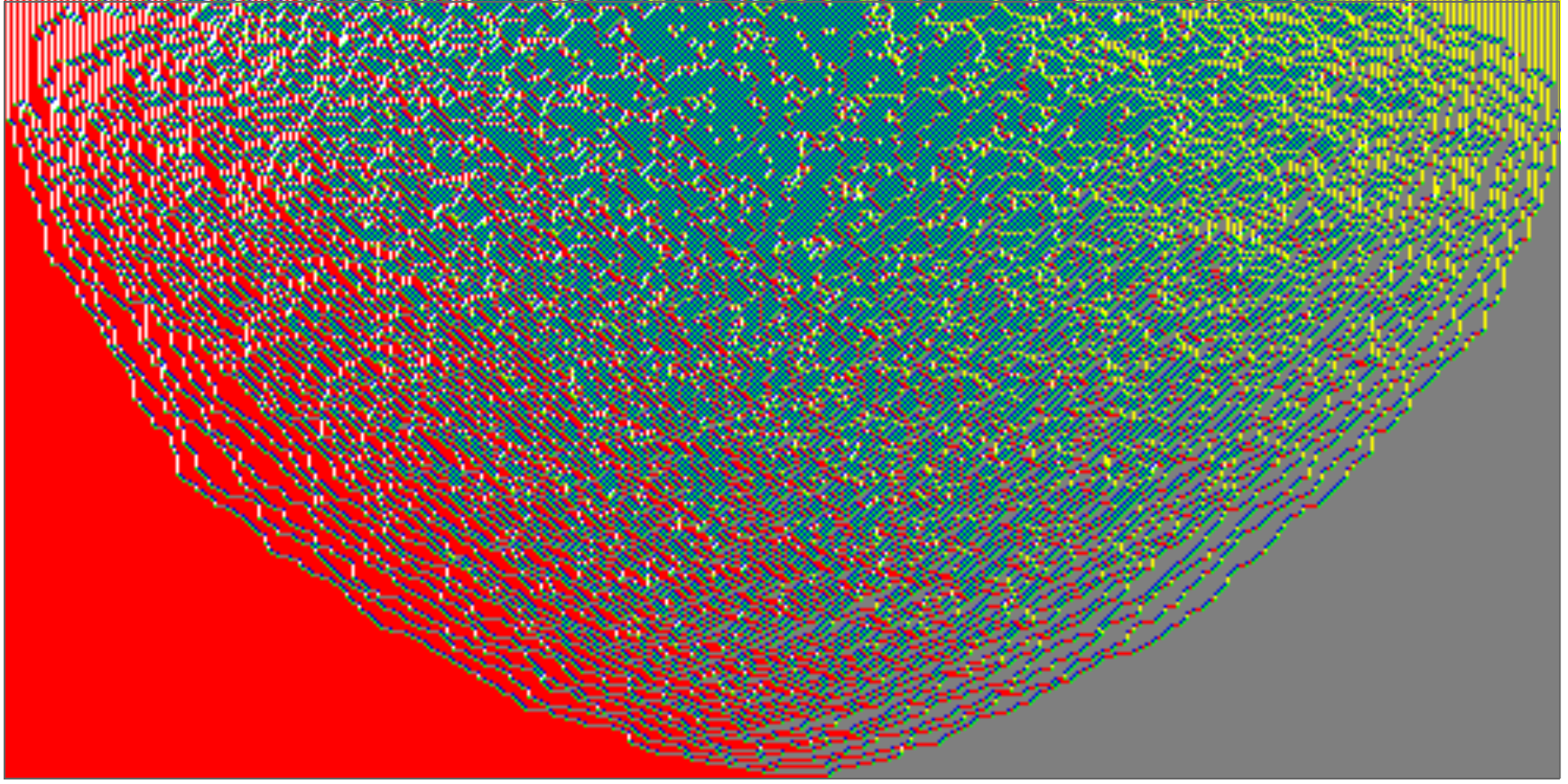}};
\node at (7.6,0.5){\includegraphics[width=0.2\textwidth]{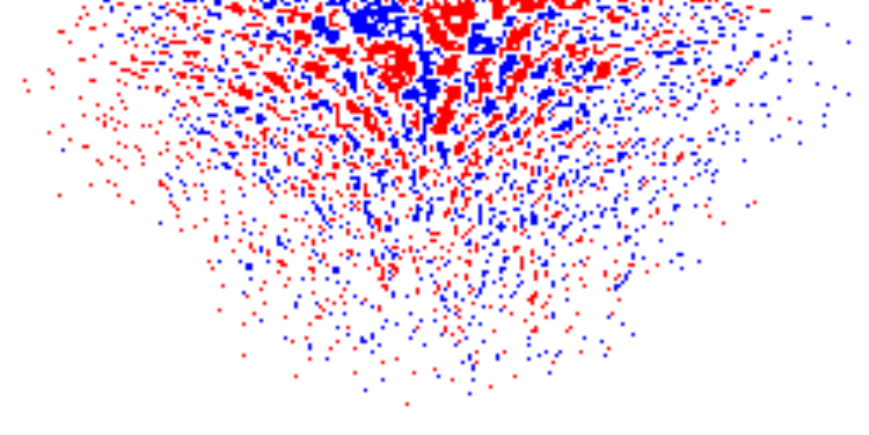}};
\node at (0,-.9){(a)};
\node at (6.8-.5,-.9){(b)};
\node at (8.7,0){(c)};
\end{tikzpicture}
\caption{(a) Six-vertex model with a RE and $b>a$, the new frozen upper left corner. (b) A typical configurations with RE at $\Delta=-2$ and $b=2a$. The vertices are coloured
as in Fig.~\ref{fig_PDWBC1}e. 
We can observe the presence of two new frozen corners at the upper boundary that are signalled by red and white stripes on the left and yellow and grey stripes on the right. 
Typical configurations contain a wide cyan region that are  antiferromagnetic phases. Inset (c). The same typical state where  antiferromagnetic patterns are coloured red and blue. We can identify an interface separating the two  phases, starting from the middle of the lattice. The interface becomes less fluctuating and more straight when $\Delta$ decreases.}
\label{frozencorners2}
\end{figure}
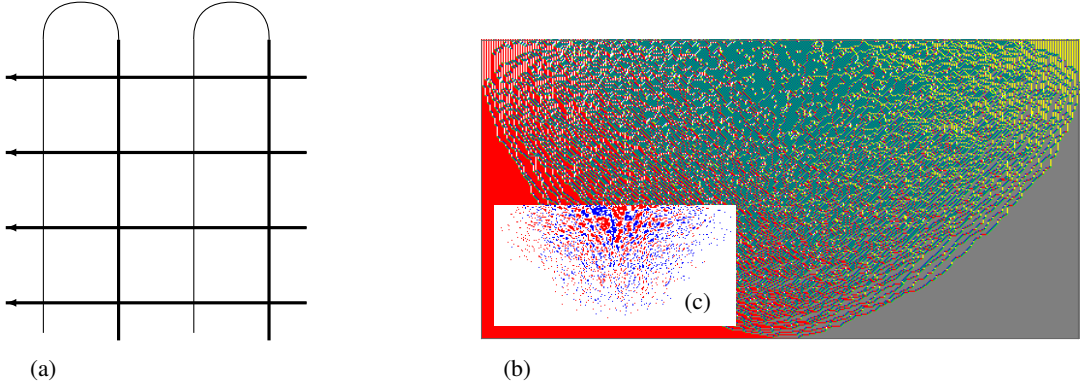

\textit{Results for $a\not=b$.~}Of course similar density plots can  be obtained for all the observables described in Sec.~\ref{sec2}.  We would like to mention here however only two facts that we believe worth to emphasize.
 \begin{itemize}
\item When $a \neq b$, the upper left and upper right corners of the rectangle may also be frozen. However the upper frozen corners have a different nature respect to the lower ones. For instance, consider the case $b>a$. 
In the upper left corner,
every vertex is of type $b_2$ on columns where the spectral parameter $\lambda$ is positive; and of type $a_1$ on columns where $\lambda$ is negative.
In the upper right corner,
every vertex is of type $b_1$ on columns where $\lambda$ is positive; and of type $a_2$ on columns where $\lambda$ is negative.
These corners are of course not as much frozen as the corners in the six-vertex model with DWBC or pDWBC. Both in the left corner and in the right corner, flips are possible on the upper boundary.
This means that there will be some impurity close to the upper boundary. However, we think that this impurity will vanish in the thermodynamic limit $N\to \infty$.
In Fig.~\ref{frozencorners2}a we show the upper left frozen corner in this case. Fig.~\ref{frozencorners2}b instead is a typical configuration with RE for $\Delta=-2$ and $b=2a$, with
vertices coloured as in Fig.~\ref{fig_PDWBC1}e. We can spot the new frozen corners on top, they are signalled by a pattern of vertical white and red stripes on the left and yellow and grey vertical stripes on the right.    
\item In the antiferromagnetic regime $\Delta<-1$, we observe the formation of antiferromagnetic domains inside the disordered phase; as discussed in Sec.~\ref{sec_pdwbc}
these consist of alternating $c_1$ and $c_2$ vertices. An example is showed in Fig.~\ref{frozencorners2}b through the appearance of a central cyan  region.
We  observe, see the inset Fig.~\ref{frozencorners2}c, that contrary to the case of DWBC, the antiferromagnetic central region  contains a fluctuating interface of vertices different from $c_1$ and $c_2$ that originates from the mid-point of the upper boundary. The curve separates two antiferromagnetic  phases. It would be interesting  to understand whether, after an appropriate rescaling, such an interface converges for $N\rightarrow\infty$ to a Brownian bridge; as for instance in the case of the two-dimensional Ising model below the critical temperature with Dobrushin boundary conditions~\cite{GI05, DV12}.
 \end{itemize}
 
\section{Six-vertex model with Half Turn Boundary Conditions}
\label{sec_ht}
\begin{figure}[t]
\centering
\begin{tikzpicture}
\draw[<->, >=latex](-0.5,0)--(5.5,0);
\draw[<->, >=latex](-0.5,1)--(5.5,1);
\draw[<->, >=latex](-0.5,2)--(5.5,2);
\draw[>-, >=latex](0,-0.4)--(0,2.5);
\draw[>-, >=latex](1,-0.4)--(1,2.5);
\draw[>-, >=latex](2,-0.4)--(2,2.5);
\draw[>-, >=latex](3,-0.4)--(3,2.5);
\draw[>-, >=latex](4,-0.4)--(4,2.5);
\draw[>-, >=latex](5,-0.4)--(5,2.5);
\draw(0,-0.5)--(0,2.5);
\draw(1,-0.5)--(1,2.5);
\draw(2,-0.5)--(2,2.5);
\draw(3,-0.5)--(3,2.5);
\draw(4,-0.5)--(4,2.5);
\draw(5,-0.5)--(5,2.5);
\draw  plot [smooth, tension=2] coordinates { (2.0,2.5) (2.5,3.0) (3.0,2.5)};
\draw  plot [smooth, tension=2] coordinates { (1.0,2.5) (2.5,3.5) (4.0,2.5)};
\draw  plot [smooth, tension=2] coordinates { (0.0,2.5) (2.5,4.0) (5.0,2.5)};
\draw(-0.5,0)--(5.5,0);
\draw(-0.5,1)--(5.5,1);
\draw(-0.5,2)--(5.5,2);
\draw(0,-0.5)--(0,2.5);
\draw(1,-0.5)--(1,2.5);
\draw(2,-0.5)--(2,2.5);
\draw(3,-0.5)--(3,2.5);
\draw(4,-0.5)--(4,2.5);
\draw(5,-0.5)--(5,2.5);
\draw[very thick](-0.5,2.0)--(2.0,2.0);
\draw[very thick](2.0,2.0)--(2.0,2.5);
\draw[very thick](-0.5,1.0)--(3.0,1.0);
\draw[very thick](3.0,1.0)--(3.0,2.0);
\draw[very thick](3.0,2.0)--(4.0,2.0);
\draw[very thick](4.0,2.0)--(4.0,2.5);
\draw[very thick](-0.5,0.0)--(4.0,0.0);
\draw[very thick](4.0,0.0)--(4.0,2.0);
\draw[very thick](4.0,2.0)--(5.0,2.0);
\draw[very thick](5.0,2.0)--(5.0,2.5);
\node at (0,-1) {(a)};
\node at (6.3,-1) {(b)};
\node at (9.8,1.5){\includegraphics[width=0.47\textwidth]{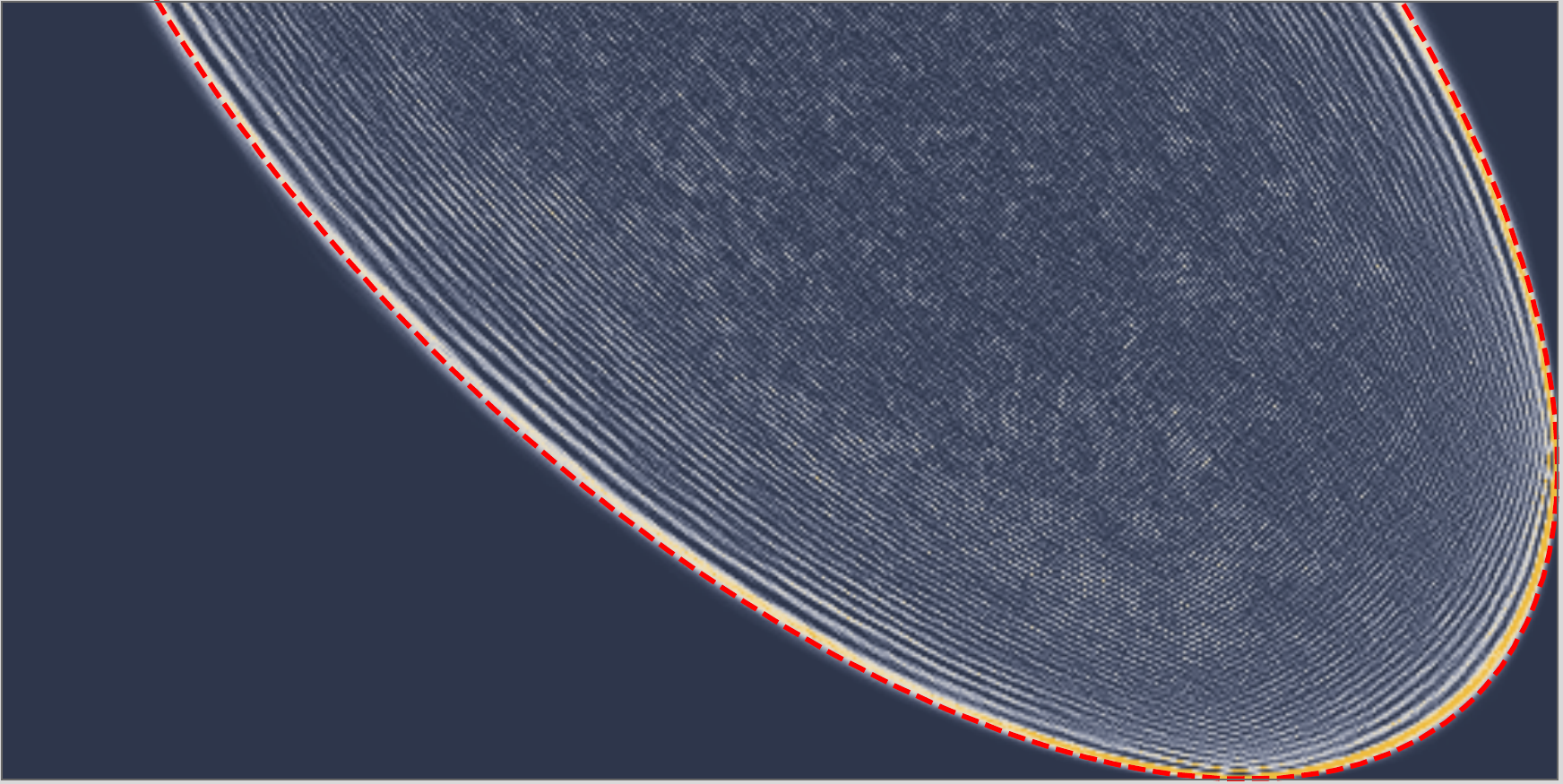}};
\node at (6.5,1){\includegraphics[height=2.4cm]{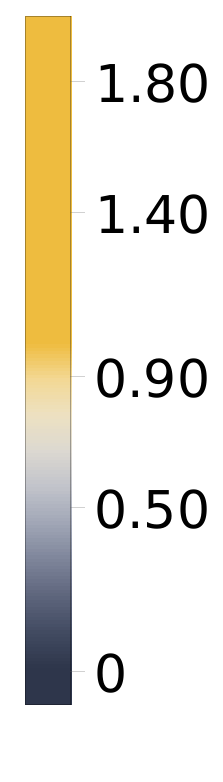}};
\node[right] at (6.8,1){$10^2\times\langle\delta\rho_c\rangle$};
\end{tikzpicture}
\caption{(a)~Six-vertex model with half turn
boundary conditions; the figure shows also an allowed state with thick curves. On the upper boundary, the vertex at $(x,y)=(2,2)$ 
can be flipped right. This vertex can also be flipped left, but only if the vertex $(4,2)$ is flipped left at the same time.
The vertex $(4,2)$ is flippable left and down. However, it can only be flipped left if the vertex $(2,2)$ is flipped left at the same time. 
(b)~$\langle\delta\rho_c\rangle$ density plot for the case $b=2a$, $c=1$ and $\Delta=0$ with half turn boundary conditions and $N=250$; 
we clearly observe the emergence of a smooth arctic curve outside of which $\langle\delta\rho_c\rangle$=0.  
The red dashed  ellipse visible in the picture is the known arctic curve~\cite{PROPP1998} in the six-vertex model with DWBC for the same parameter values (of course when $N\rightarrow\infty$).}
\label{fig_htbc1}
\end{figure}
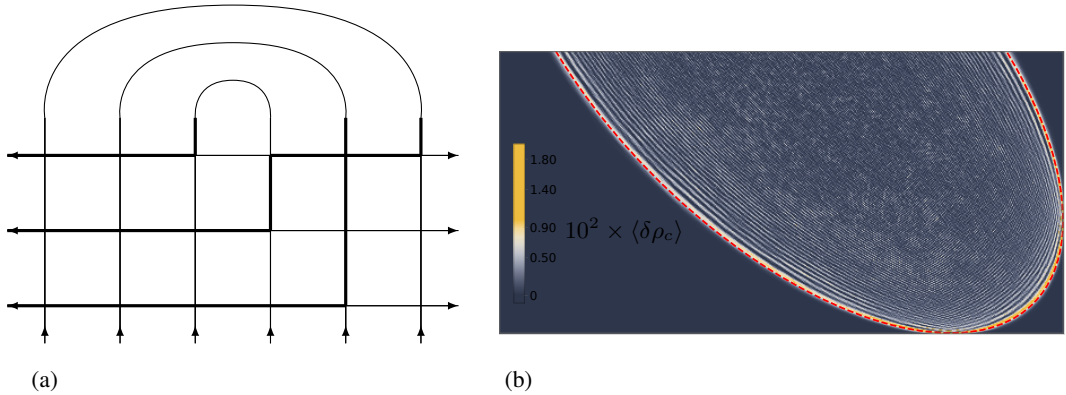
\textit{Model and algorithm.~}The six-vertex model with half turn boundary conditions (HTBC)
is defined on a rectangular lattice of size $2N\times N$ \cite{BL2017}. The boundary conditions are defined as on Fig. \ref{fig_htbc1}a. 
Arrow configurations of the six-vertex model with HTBC on a $2N\times N$ rectangle are in one-to-one correspondence with arrow configurations of the six-vertex model with DWBC on $2N\times 2N$ square that  are invariant under a $180^{\text{o}}$ rotation. The partition functions~\cite{BL2017} are also simply related if one assumes $w(a_1)=w(a_2)=a$, $w(b_1)=w(b_2)=b$, $w(c_1)=w(c_2)=1$ as we will also do in the following. However, notice that as for pDWBC the six-vertex model with HTBC would have in principle an extra parameter defined  from $w(a_1)/w(a_2)=w(b_1)/w(b_2)= e^{-2\eta}$.

With HTBC there are the same seven flips as in section \ref{sec_pdwbc}.
However, it is usually not possible to perform only one left or right flip.
Unless $x = N$, a left flip at $(x,N-1)$ has to happen together with a left flip at $(2N-x,N-1)$. Similarly, 
unless $x = N - 1$, a right flip at $(x,N-1)$ has to happen together with a right flip at $(2N-x-2,N-1)$.  

In order to implement this rule, we define a second set of lists, in addition to the seven lists defined in sections \ref{sec2} and \ref{sec_pdwbc}. We call the new lists
$L_u'$, ..., $L_{ld}'$. A point $p=(x,N-1)$ can only belong to one of the lists $L_{l}'$ and $L_{ld}'$ if $x\leq N$. It can 
only belong to one of the lists $L_{r}'$ and $L_{lr}'$ if $x\leq N - 1$. The reason why this restriction is used is that we want to avoid double counting. Now suppose that
$x\leq N$. Let $p=(x,N-1)$, $p$ is flippable left (and only left) precisely when it belongs to the list $L_l'$. 
We define the two related points $q_l=(2N-x,N-1)$ and $q_r=(2N-x-2,N-1)$. Then 
\be
p\in L_l' \Leftrightarrow \Bigl( p\in L_l ~\wedge ~  q_l \in \bigl(L_l \cup L_{lr} \cup L_{ld}\bigr) \Bigr) \vee  
\Big( p\in L_{lr} ~\wedge ~  q_l \in\bigl(L_l \cup L_{lr} \cup L_{ld}\bigr) ~\wedge ~  q_r \notin\bigl( L_r \cup L_{lr}\bigr)  \Big).
\label{leftlist}
\ee
The other six new lists are defined similarly.

We will describe how flips on the upper boundary are performed.
The left flip at a vertex $v=(x,N-1)$ which is left flippable only is performed as follows.
If $x=N$, then the left flip can be done freely. As before, we call the state of pair of vertices $\{(N-1,N-1),(N,N-1)\}$ before the
flip $T$, and the state after the flip $T'$.
The probability of the left flip is then, as before, 
\be P_{l} = \frac{Z_v(T')}{R}.
\label{ht1}
\ee
If $x\neq N$, then the flip can only happen if the vertex at $(2N-x,N-1)$ can also be flipped left. 
We use the notation $v_1\equiv(x,N-1)$ and $v_2\equiv(2N-x,N-1)$.
The probability of a left flip at $v_1$ is thus the same as the probability of the
simultaneous left flips at $v_1$ and $v_2$;
\be  P_{l} = \frac{Z_{v_1}(T_1')Z_{v_2}(T_2')}{R},
\label{ht2}
\ee
where $T_1'$ is the state after a left flip at the vertex $v_1$ and $T_2'$ the state after a left flip at $v_2$.

We will also describe flips at a vertex which is left and right flippable. 
There are three cases to consider. The first case is $(x,y)=(N-1,N-1)$. We call this vertex $v_1$. We call the state after a left flip $T_L'$, and the state
after a right flip $T_R'$. If the flip is to the left, then there will also be a left flip at the vertex $v_2$ at $(x,y)=(N+1,N-1)$. We call the state
after the left flip at $v_2$, $U'$. There will be a flip with probability 
\be P_{lr} = \frac{Z_{v_1}(T'_L)Z_{v_2}(U')+Z_{v_1}(T'_R)}{R}.
\label{ht3}
\ee
If it is decided that there will be a flip, then the flip will left with probability $ P_{l|lr}$  and right with probability $P_{r|lr}$, where
\be P_{l|lr} = \frac{Z_{v_1}(T'_L)Z_{v_2}(U')}{Z_{v_1}(T'_L)Z_{v_2}(U')+Z_{v_1}(T'_R)},\quad P_{r|lr} = \frac{Z_{v_1}(T'_R)}{Z_{v_1}(T'_L)Z_{v_2}(U')+Z_{v_1}(T'_R)}.
\label{ht4}
\ee
The second case is $(x,y)=(N,N-1)$; since it is similar to the previous case, it will be omitted. The third case is $x\neq N-1,~N$. 
We use the same notation as in the first case. 
If the flip is to the left, then there will also be a left flip at the vertex $v_2$ at $(2N-x,N-1)$.
If the flip is to the right, then there will also be a right flip at the vertex $v_3$ at $(2N-x-2,N-1)$.
We call the state after the left flip at $v_3$, $V'$. There will be a flip with probability 
\be P_{lr} = \frac{Z_{v_1}(T'_L)Z_{v_2}(U')+Z_{v_1}(T'_R)Z_{v_3}(V')}{R}.
\label{ht6}
\ee
If it is decided that there will be a flip, then the flip will left with probability $ P'_{l|lr}$ and right with probability $P'_{r|lr}$ where now
\be P'_{l|lr} = \frac{Z_{v_1}(T'_L)Z_{v_2}(U')}{Z_{v_1}(T'_L)Z_{v_2}(U')+Z_{v_1}(T'_R)Z_{v_3}(V')},\quad P'_{r|lr} = \frac{Z_{v_1}(T'_R)Z_{v_3}(V')}{Z_{v_1}(T'_L)Z_{v_2}(U')+Z_{v_1}(T'_R)Z_{v_3}(V')}.
\label{ht7}
\ee
We tested the above algorithm against exact enumeration obtaining for instance for a $8\times 4$
 rectangle $\langle E\rangle_{\text{exact}}=0.23346$ and $\langle E\rangle=0.23352\pm 0.00060$; the two results are compatible inside the error bar.\newline\newline
\begin{figure}
\centering
\begin{tikzpicture}
\node at (0,0) {\includegraphics[height=4.5cm]{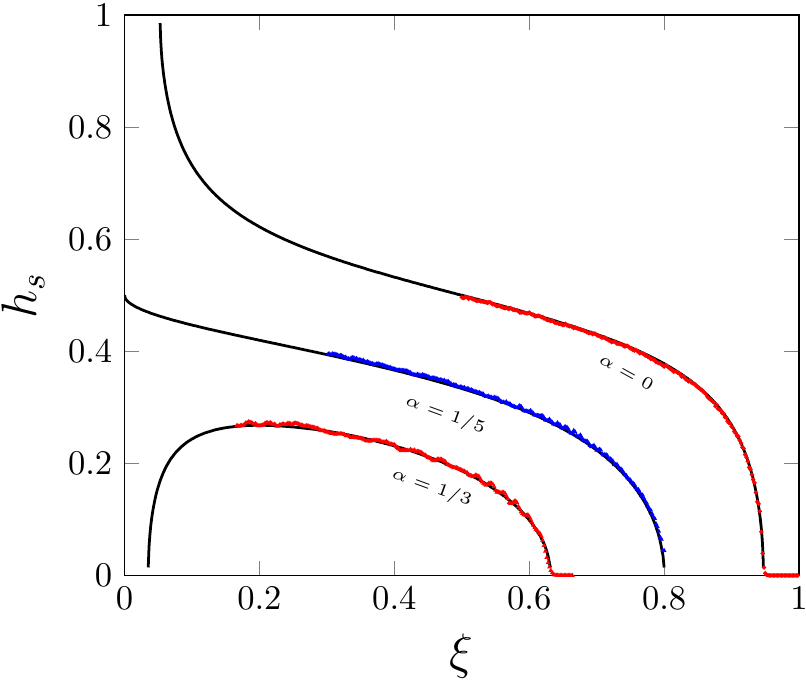}};
\node at (8.5,0.3){\includegraphics[width=0.47\textwidth]{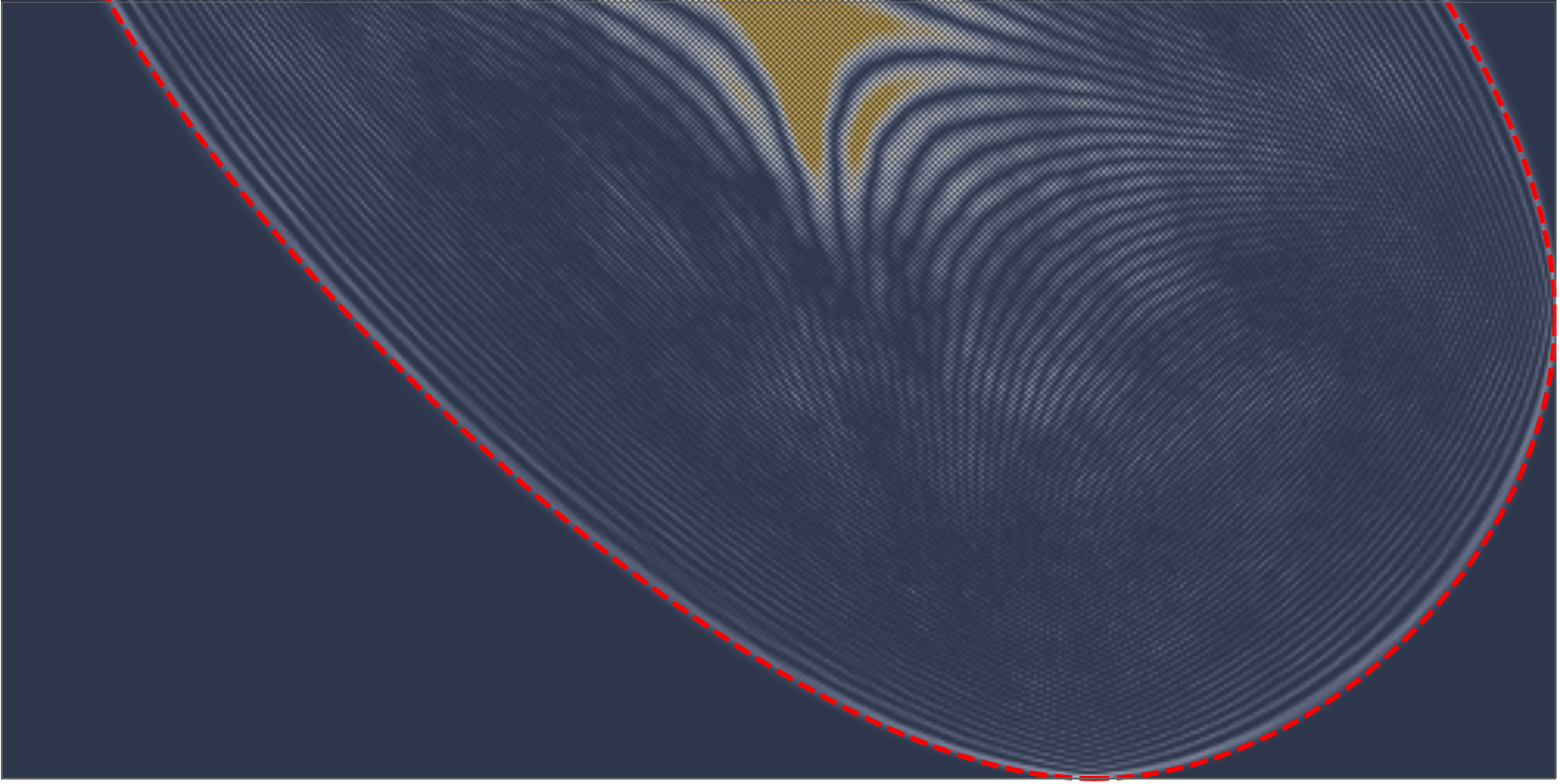}};
\node at (5.5,0){\includegraphics[height=2.3cm]{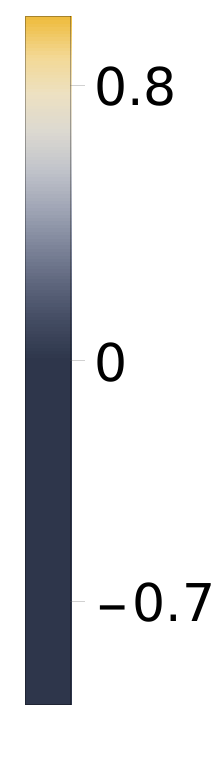}};
\node at (6.5,-0.4) {$20\times\langle\delta\rho_c\rangle$};
\node at (-1,-2.3){(a)};
\node at (6,-2.3){(b)};
\end{tikzpicture}
\caption{(a)Six-vertex model with HTBC at $\Delta=0$ and $b=2a$. The vertical edge density estimated with MC is compared with the exact result (black curves) given in \eqref{asympt} for $b=2a$. The agreement is satisfactory already when $N=250$. (b) The density plot of $\langle\delta\rho_c\rangle$ for $\Delta=-2$ and $b=2a$, the red dashed line is the known arctic curve~\cite{PKZJ} for DWBC with the same values of the parameters.}
\label{fig_htbc_2}
\end{figure}
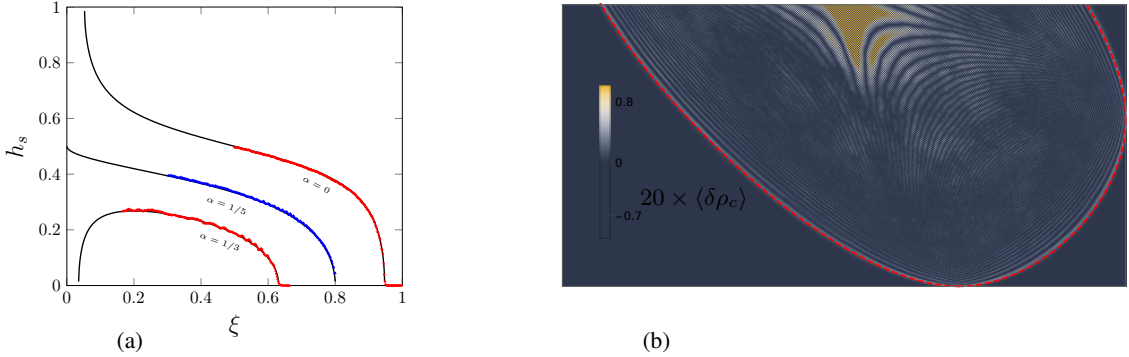
\textit{The arctic curves and the density profiles.~} We consider $\eta=0$ for simplicity. We will present results for two representative cases $\Delta=0$ and $b=2a$ in
the  disordered regime and $\Delta=-2$, $b=2a$ in the antiferromagentic regime.  Fig.~\ref{fig_htbc1}b shows the density
plot of $\langle\delta\rho_c\rangle$, at $\Delta=0$, $b=2a$ for $N=250$. We can identify the arctic curve, separating a frozen region with
$\langle\delta\rho_c\rangle=0$ from a domain where the same quantity is non-vanishing. The arctic curve coincides with the lower half of the arctic curve for
DWBC (arctic ellipse,~\cite{PROPP1998}), for the same values of the parameters. The same is true for $\Delta =-2$ and $b=2a$ as it can been clearly seen from Fig.~\ref{fig_htbc_2}b.
In hindsight, this is perhaps not surprising, in~\cite{BL2017} for instance it is proved that  the extensive part of the the free-energy is the same for the two models in the thermodynamic limit.

Finally, let us now denote by $h_s(x)$ the vertical edge density computed on the straight line $r_s$ of Fig.~\ref{rdwbcfig}b on a $2N\times N$ lattice with HTBC, i.e. $h_s(x)=\langle\rho_v(x,y)\rangle|^{\text{HTBC}}_{(x,y)\in r_s}$.
Similarly, we also found that at $\Delta=0$, $h_s(x)$  converges in the thermodynamic limit to the same quantity calculated with DWBC, see Sec.~\ref{rdwbc} for more details.
A numerical verification of the last statement is presented in Fig.~\ref{fig_htbc_2}a, where the function $h_s(2N\xi)$ obtained from the MC is compared for different values of $s=2N\alpha$ with the conjectured exact asymptotic \eqref{asympt} at $b=2a$. The error bar is not visible on a such a scale (see also Appendix~\ref{apb}). The agreement is satisfactory already with $N=250$ and a similar accuracy is obtained varying the parameters $a$ and $b$, when $\Delta=0$. At present, no analytical derivation  of this result is available.
\section{Conclusions}
\label{sec_con}
In this paper we analyzed  numerically phase separation in the six-vertex model with a variety of boundary conditions.
We modified the original AR algorithm, to address the study of pDWBC, the six-vertex model with a RE and HTBC.
In some cases~\cite{BL2014, RK2015, BL2017}, exact results for the partition functions were known in the thermodynamic limit but the
presence of phase separation and the shape of the arctic curve have been never investigated. We show that when the model has only two parameters $a$ and $b$,
typical configurations display features similar to DWBC, varying $\Delta$ in Eq.~\eqref{delta}. Summarizing we have found strong numerical evidence that
\begin{itemize}
\item The six-vertex model with pDWBC and $\eta=0$ (see Sec.~\ref{sec_pdwbc}) displays phase separation and there exist frozen corners of the same type as with DWBC. In the thermodynamic limit, typical
configurations
for $\Delta<1$ should contain a piecewise differentiable arctic curve. It would be of clear interest to determine analytically the limiting shapes. 
\item The six-vertex model with one reflecting end and symmetric $U$-turn weights (see Sec.~\ref{rdwbc}) displays phase separation and there exist frozen corners.
If in particular $a=b$, in all the examples we
investigated ($\Delta=0$, $\Delta=-2$) we obtained that the
arctic curve in the thermodynamic limit converges to the lower half of the exact curve~\cite{PC2010, PKZJ} derived for the model with DWBC. This is consistent with the recent findings in~\cite{DF2017} for vertically symmetric alternating sign matrices (corresponding to $a=b$ and $\Delta=1/2$). Moreover at $\Delta=0$ the same is true for the
vertical edge density profile. If $a\not=b$ (in particular $b=2a$), in the model with one RE we observed the formation of new frozen corners and a clear qualitative difference in the
shape of the arctic curve in the thermodynamic limit compared with DWBC. Again it would be important to confirm analytically this numerical observation.
\item Arctic curves in the six-vertex model with HTBC and $\eta=0$ (DWBC with half turn symmetry~\cite{BL2017}) are found to be, in the thermodynamic limit,  the lower half of the
known arctic curves~\cite{PC2010, PKZJ} with DWBC, in all the cases we examined ($\Delta=0,~\Delta=-2,~b=a,~b=2a$). The same is true at $\Delta=0$ for the vertical edge density
profile. Numerics seems to indicate that in the thermodynamic limit, such a model is equivalent to the lower half of the six-vertex model with DWBC.
\end{itemize}
 
As a future perspective, it would be natural to investigate what is the  more general class of boundary conditions that lead to the same arctic curves in the thermodynamic limit as DWBC. 
Finally, the Metropolis algorithm could be also adapted to the nineteen vertex model, discussed for instance  in~\cite{E2017}.   
\section*{Acknowledgements} We thank F. Colomo and  J. Lamers for a critical reading of the manuscript and for valuable comments. We are also grateful to P. Bleher for correspondence. JV thanks the ICTP/SAIFR of S\~ao Paulo for hospitality.
\appendix
\section{DWBC with $\Delta=-2$}
\label{apa}
\begin{figure}[t]
\centering
\begin{tikzpicture}
\node at (0,0) {\includegraphics[height=6cm]{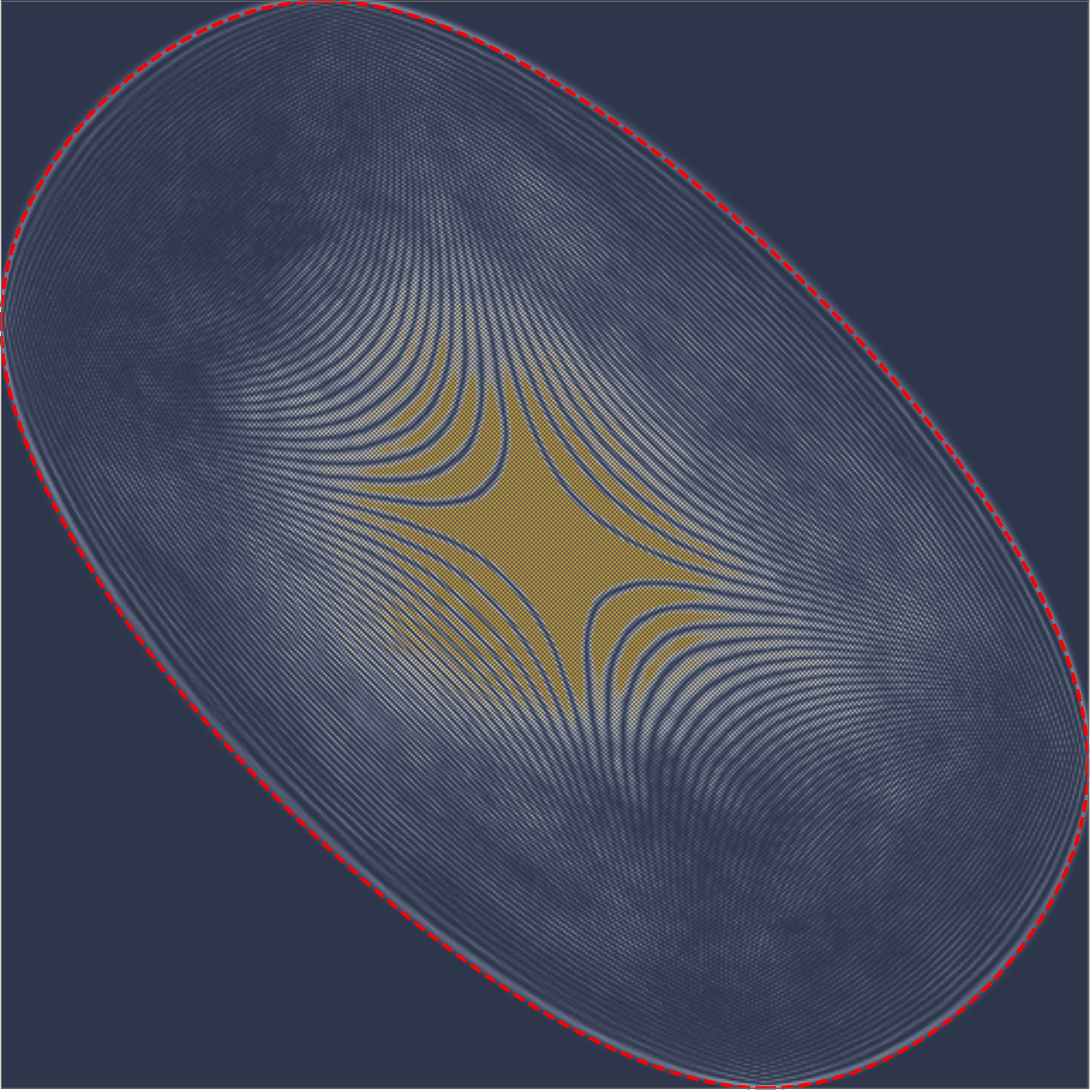}};
\node at (8,-1.5) {\includegraphics[height=3cm]{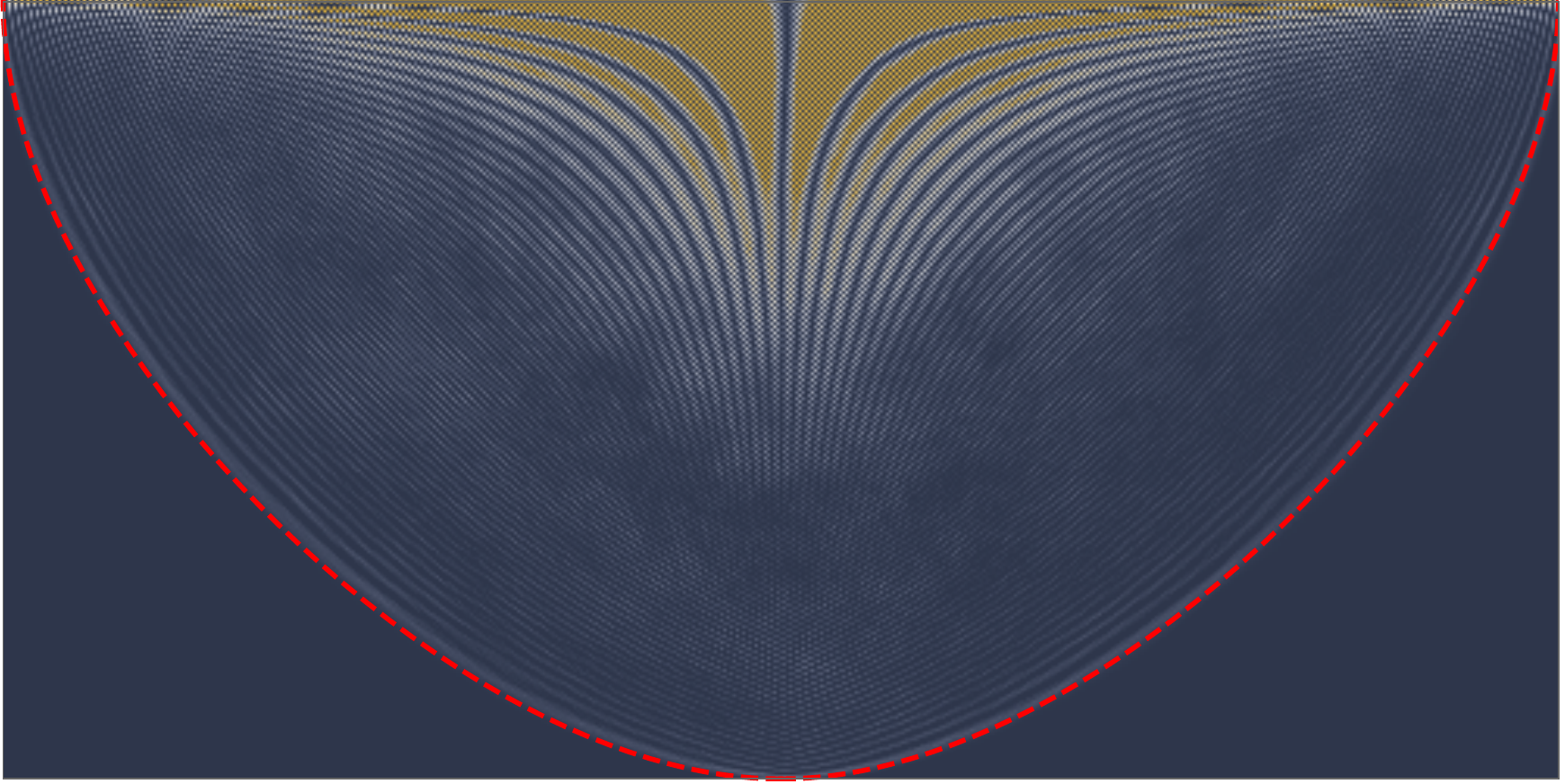}};
\node at (-2.5,-1.8){\includegraphics[height=2.3cm]{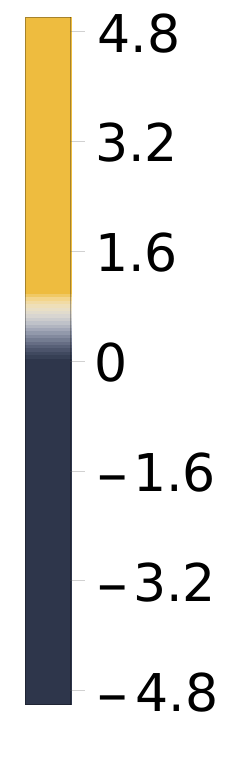}};
\node at (6,-1.2){\includegraphics[height=2.3cm]{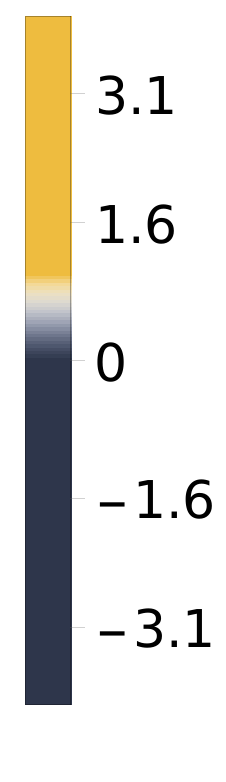}};
\node at (-1.4,-2.3){$10\times\langle\delta\rho_c\rangle$};
\node at (-1.4+8.5,-1.6){$10\times\langle\delta\rho_c\rangle$};
\node at (-2.5,-3.2) {(a)};
\node at (-2.5+8,-3.2) {(b)};
\end{tikzpicture}
\caption{(a) Density plot $\langle\delta\rho_c\rangle$ for DWBC with $N=500$, $\Delta=-2$, $b=2a$. The red dashed line is the exact (in the thermodynamic limit) arctic curve~\cite{PKZJ}. (b) Density plot $\langle\delta\rho_c\rangle$ for the six-vertex model with a RE with $N=250$, $\Delta=-2$, $b=a$. The red dashed line is the lower half of the exact (in the thermodynamic limit) arctic curve~\cite{PKZJ} calculated for DWBC with the same values of the parameters.}
\label{fig_dwbcm2}
\end{figure}
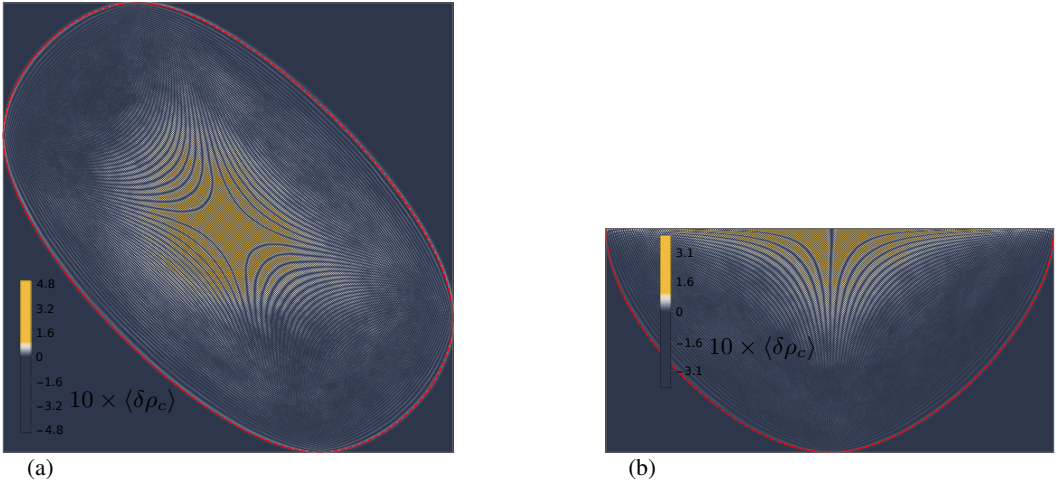
We have mentioned in the Sec.~\ref{intro} and Sec.~\ref{sec2} that the thermalization time of the Metropolis algorithm  increases in the antiferromagnetic regime, especially when $b\not= a$. For instance on a square lattice $500\times 500$ with DWBC at $\Delta=-5$ and $b=2a$ we were not able to recover the exact arctic curve after more than 48h of simulation on a desktop computer. This is  also evident looking at the corresponding images contained in~\cite{AR2005}. We therefore  consider as a lower bound for our analysis $\Delta=-2$, where also for $b=2a$ we could recover the exact limiting shape with DWBC. In particular, the density plot $\langle\delta\rho_c\rangle$ with DWBC for $N=500$, $\Delta=-2$ and $b=2a$ is reported in Fig.~\ref{fig_dwbcm2}a where also the exact thermodynamic limit of the arctic curve~\cite{PKZJ} is drawn as a red dashed line. 
\section{Convergence to the thermodynamic limit of the density profiles}
\label{apb}
\begin{figure}[t]
\centering
\begin{tikzpicture}
\node at (0,0) {\includegraphics[height=5cm]{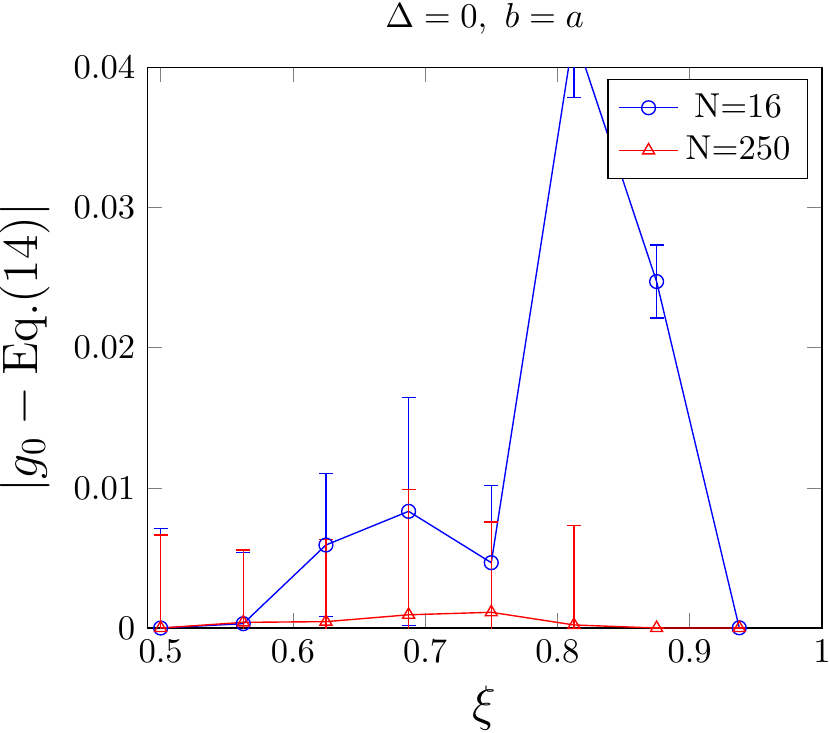}};
\node at (8,0) {\includegraphics[height=5cm]{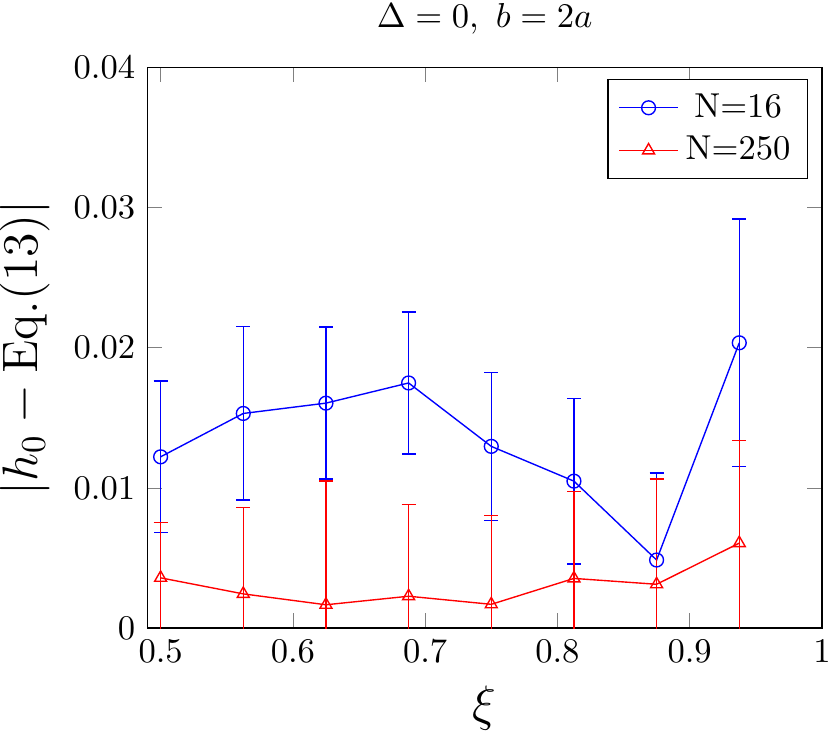}};
\node at (-1,-2.5) {(a)};
\node at (-1+8,-2.5) {(b)};
\end{tikzpicture}
\caption{(a)~Six-vertex model with a RE at $\Delta=0$ and $a=b$. Absolute value of the difference between $g_0(2N\xi)$ (see Sec.~\ref{rdwbc} for a definition) and the right hand side of Eq.~\ref{conj} for $\alpha=0$. The data sets correspond to $N=16$ and $N=250$. (b)~Six-vertex model with HTBC (DWBC with half turn symmetry,~\cite{BL2017})  at $\Delta=0$ and $b=2a$. Absolute value of the difference between $h_0(2N\xi)$ (see Sec.~\ref{sec_ht} for a definition) and the right hand side of Eq.~\eqref{asympt} for $\alpha=0$ and $b=2a$. The data sets correspond to $N=16$ and $N=250$.}
\label{fig_error}
\end{figure}
In Sec.~\ref{rdwbc} and Sec.~\ref{sec_ht} we have numerically observed that the vertical edge density profiles at $\Delta=0$ for the six-vertex model with one RE and with HTBC converge, in the thermodynamic limit $N\rightarrow\infty$, to the same quantity calculated with DWBC, see Eq.~\eqref{asympt}. The statement holds only at $a=b$ for RE and for any allowed $a$ and $b$ for HTBC. The figures Fig.~\ref{vertical}a and Fig.~\ref{fig_htbc_2}a were numerical evidences  obtained with our MC algorithm. In this Appendix we focus on the case $s=0$ and present a more detailed analysis of the convergence of the vertical edge density profiles to their conjectured expression for $N\rightarrow\infty$. In Fig.~\ref{fig_error}a we plot the absolute value of the difference between $g_0(2N\xi)$ and the right-hand-side of Eq.~\eqref{conj} for $\alpha=0$, varying $N$. In particular for $N=16$ and $N=250$, the last value of $N$ corresponding to the plot in Fig.~\ref{vertical}a. Analogously in Fig.~\ref{fig_error}b we plot the absolute value of the difference between $h_0(2N\xi)$ and its conjectured thermodynamic limit Eq.~\eqref{asympt} for $\alpha=0$, $b=2a$ and again $N=16$ and $N=250$. We notice that in the second case convergence appears slower, this might be related to the fact that the thermalization time increases when $b\not=a$.

\end{document}